\documentclass[a4paper]{jpconf}
\RequirePackage{doi}
\usepackage{graphicx}
\usepackage{xcolor}
\usepackage{bm}
\usepackage{float}
\usepackage{hyperref}

\newcommand{\mgf}{{\cal B}}
\newcommand{\MGF}{\vec{\cal B}}
\newcommand{\AVP}{\vec{\cal A}}

\newcommand{\mbfp}{{\mathbf{p}}}
\newcommand{\mbfr}{{\mathbf{r}}}

\newcommand{\HerH}{{\rm H}}

\newcommand{\LerchL}{{\rm L}}
\newcommand{\sgn}{{\rm sgn}}

\begin{document}

\title{Transient capture of electrons in magnetic fields, or: comets in the restricted three-body problem}

\author{Tobias Kramer}
\address{Zuse Institute Berlin, Supercomputing Department, Berlin, Germany}
\address{Harvard University, Department of Physics, Cambridge, MA~02138, USA}
\ead{kramer@zib.de}

\date{\today, Version 0.1}

\begin{abstract}
The motion of celestial bodies in astronomy is closely related to the orbits of electrons encircling an atomic nucleus.
Bohr and Sommerfeld presented a quantization scheme of the classical orbits to analyze the eigenstates of the hydrogen atom.
Here we discuss another close connection of classical trajectories and quantum mechanical states: the transient dynamics of objects around a nucleus.
In this setup a comet (or an electron) is trapped for a while in the vicinity of an parent object (Jupiter or an atomic nucleus), but eventually escapes after many revolutions around the center of attraction.
\end{abstract}

\section{Introduction: celestial and quantum mechanics}

The semiclassical method for quantization provides the link between classical and quantum mechanics and highlights the role of closed orbits for the determination of eigenvalues and eigenfunctions.
For systems where the electrons are not bound photodetachment and photoionization processes are amendable to a semiclassical treatment \cite{Fabrikant1980,Demkov1982,Peters1993,Bordas1998,Kramer2001,Bracher2005,Kanellopoulos2009,Kramer2011a}.
A spectacular experimental realization of the macroscopic extension of the photodetachment wavefunction was shown by Blondel et al \cite{Blondel1996,Blondel1999}.
The photodetachment ``microscope'' provides the most accurate values of the electron affinities.

{\color{black}Additional molecular states are amendable to a (semi)classical interpretation.
The theoretical prediction \cite{Greene2000} and later experimental realization \cite{Bendkowsky2009} of highly excited Rydberg molecules mirrors directly the underlying Keplerian orbits \cite{Granger2001}.
Making use of Lambert's theorem for cometary motion \cite{Lambert1761}, the bound and unbound eigenstates and energies of the quantum mechanical Coulomb are accessible with semiclassical methods \cite{Kanellopoulos2009}}.

{\color{black}By solving the classical equations of motions for electrons in external potentials, or the time-dependent Schr\"odinger equation for a quantum system, a physical picture emerges which emphasizes the dynamics rather than the stationary eigenstates \cite{Tannor2008,Kramer2011a,Heller2018}.
From the computational point of view, this offers the possibility to efficiently solve an initial value problem by propagating electrons rather than a numerically demanding boundary value problem.
}

{\color{black}Here, we discuss two topics with similar equations of motion and phase space structures, both amendable to very similar numerical techniques.}
The classical equations of motions for certain restricted three-body problems of celestial mechanics resemble closely the equations for the motion of an electron around a positively charged nucleus within a magnetic field.
Besides in atomic and molecular physics, the dynamics of electrons in external magnetic fields also plays an important role in nano-structures, where electrons follow cyclotron orbits or can be confined inside a quantum dot \cite{Aidala2007,Kramer2008}.
{\color{black}The Hamiltonian discussed in Sect.~\ref{sec:ClassicalHamiltionian} is the workhorse for studying the experimentally observed Aharonov-Bohm oscillations in nano-structured wave-guides embedded within a two-dimensional heterostructure \cite{Kramer2010,Kramer2016AB,Kreisbeck2017}}.

The electron dynamics in a wave guide within a magnetic field bears a close correspondence to celestial mechanics, further explored in Sect.~\ref{sec:3body}.
{\color{black}Both systems feature a mix of bound and continuum states depending on the initial conditions.}
In celestial mechanics, a common question is the stability of cometary orbits perturbed by the planet Jupiter.
Jupiter sometimes captures comets which are then encircling Jupiter rather than the sun \cite{Benner1995}.
These trapped comets will eventually escape or might impact Jupiter -- as witnessed also by the author in July 1994 when comet Shoemaker-Levy 9 (SL9) crashed into Jupiter and produced large disturbances in its atmosphere.
The comet-capture process can be considered as a restricted three-body problem, where Jupiter's gravity is the main center of attraction, while the solar attraction is incorporated by the transformation to a co-rotating frame with Jupiter.

The general interacting many-body problem is challenging for both, classical and quantum mechanics, with closed form solutions available for two interacting electrons \cite{Vercin1991,Kramer2010a,Grossmann2011}
For several interacting electrons in a magnetic field quantum mechanical computations require a suitable adapted basis expansion that avoids spurious states, see \cite{Kramer2015} for the analysis of the four-electron case.

\section{Electrons in a homogeneous magnetic field and a parabolic wave-guide.}\label{sec:ClassicalHamiltionian}

The quantum theory of electrons in external potentials and magnetic fields shows a wealth of unexpected features seen in experiments. 
For large magnetic fields, a quantized Hall conductivity occurs in a two-dimensional electron gas, which is subject to a perpendicular magnetic field. In weaker magnetic fields, the electrons move along bouncing-ball cyclotron orbits along the confining gates \cite{VanHouten1989,Aidala2007}.
We discuss first the analytic solvable model of electron motion confined by a potential barrier, i.e.\ the potentials generated by a gated structure or the sample edges. 
One application of this model is the theory of the quantum Hall effect and the determination of the density of electrons within each Landau level (the filling factor).
The filling factor changes with the addition of a curved potential and also the velocity of the drifting electrons is affected.
The product of both quantities constitutes the drift current and it has been sometimes conjectured that both quantities change in an exactly reciprocal way to keep the current itself invariant.
This would preserve a conductivity exactly quantized in units of $e^2/h$\cite{Aoki1981,Prange1981,Uemura1983,Brenig1983}. 
However, this ``plausible'' \cite{Brenig1983} conjecture was only investigated for impurities distributed on a flat potential.
The assumption of a constant potential is a severe constraint and does not account for long-range disorder potentials or smoothly varying confining potential.
The parabolic waveguide provides a counter-example to the assumption of a reciprocal change of LDOS and drift velocity. 
Previous work on this system has focused on the global density of states (DOS) (see references in Ref.\ \cite{Beenakker1991}, Sect.\ 12) and did not analyze the details of the local current flow.

\subsection{Hamiltonian and classical equations of motion}

\begin{figure}[t]
\begin{center}
\includegraphics[width=0.49\textwidth]{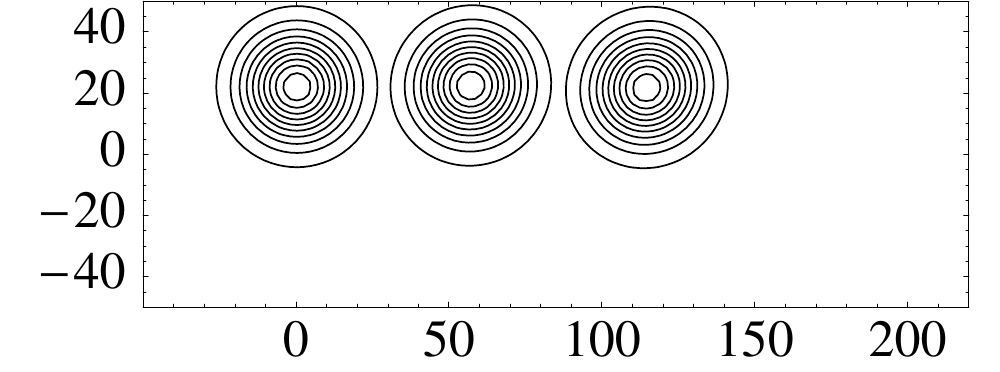}\hfill
\includegraphics[width=0.49\textwidth]{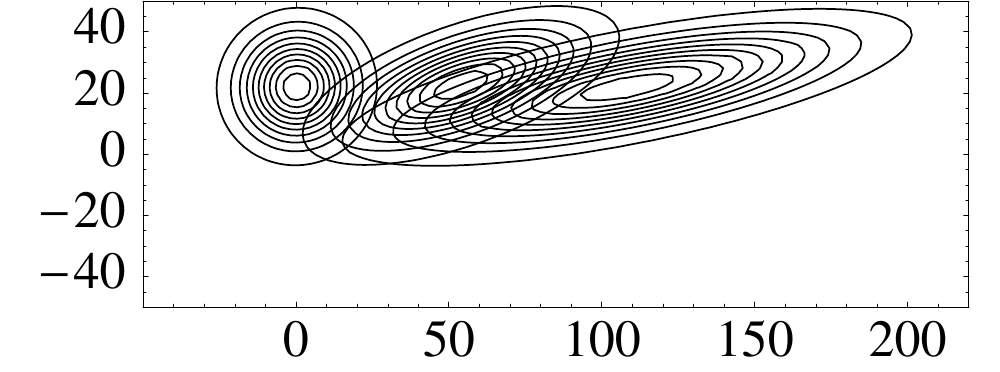}
\end{center}
\caption{
Snapshots of an initially Gaussian wavepacket at three different times. The left panel depicts the evolution in a linear potential (0 curvature), while the right panel shows the evolution for a curvature $A=0.4$. The two centers of the wavepacket drift with almost exactly the same velocity.
However, in the presence of a curvature the wavepacket is also spreading out against the drift direction. The spreading leads to a larger overlap with the initial wavepacket and marks a very weak localization effect, which manifests in the singularities of the LDOS. Parameters: 
${\cal B}=5$~T, 
${\cal E}=4000$~V/m, 
$A=0.4$, 
$\Omega=\omega_L\sqrt{4+A^2}\approx 2.04\;\omega_L$, 
$x=0, y=\frac{e{\cal E}}{m A^2 \omega_L^2}\approx 22.7$~nm.
\label{fig:wp}}
\end{figure}
We consider an electron released in a homogeneous magnetic field ${\cal B}$ aligned along the $z$-axis, and a parabolic potential $V(y)=\frac{1}{2}m\omega_y^2 y^2$ along the $y$-direction. We assume a strongly confining potential along the $z$-axis, which quantizes the system along the $z$-direction and thus reduces the dimensionality of the subsystem under consideration to two.  
{\color{black}This configuration is commonly realized experimentally in semiconductor heterostructures.}
The Hamiltonian for the two-dimensional motion is conveniently expressed in the Landau gauge $\AVP=(-{\cal B} y,0,0)$ \cite{Johnson1983}:
\begin{equation}\label{eq:HamiltonParab}
H=\frac{1}{2m}{(-\rmi\hbar\partial_x+q{\cal B}y)}^2-\frac{\hbar^2}{2m}\partial_y^2+\frac{1}{2}m \omega_y^2 y^2.
\end{equation}
For easier notation, it is useful to introduce the Larmor frequency $\omega_L$ and to express $\omega_y$ in terms of another frequency $\Omega$
\begin{equation}
\omega_L=\frac{e\mgf}{2m},
\quad
\Omega^2=\omega_y^2+4\omega_L^2.
\end{equation}
The Hamiltonian is a quadratic polynomial in $\mbfr, \mbfp$, therefore one can readily derive the classical action and the equations of motions:
\begin{eqnarray}
\mbfr(t)&=&
\left(
\begin{array}{cc}
-\frac{2\omega_L p_y(0)}{m\Omega^2} & \frac{2\omega_L(2p_x(0)\omega_L+my(0)(\Omega^2-4\omega_L^2))}{m\Omega^3}\\
y(0)+\frac{2\omega_L p_y(0)}{m\Omega^2}-\frac{4 y(0)\omega_L^2}{\Omega^2} & \frac{p_y(0)}{m\Omega}\\
\end{array}
\right)
\left[\begin{array}{c}
\cos(\Omega t)\\
\sin(\Omega t)
\end{array}\right]\nonumber\\
&+&
\left[
\begin{array}{r}
\frac{(p_x(0)-2my(0)\omega_L)(\Omega^2-4\omega_L^2)}{m\Omega^2} t 
+ \frac{m x(0)\Omega^2+2p_y(0)\omega_L}{m\Omega^2}\\
-\frac{2\omega_L(p_x(0)-2my(0)\omega_L)}{m\Omega^2}
\end{array}
\right].
\end{eqnarray}
The linear term in $t$ in the second $[]$-brackets describes a uniform drift motion that depends on the initial position $y(0)$ and momentum $p_x(0)$. However, the angular averaged drift velocity for a given energy $E$ is independent of the energy:
\begin{equation}
\langle v_d \rangle_{cl}=\frac{1}{2\pi}\int_0^{2\pi}
\rmd \alpha \frac{(\sqrt{2 m E}\cos\alpha-2 my(0)\omega_L)(\Omega^2-4\omega_L^2)}{m\Omega^2}
=2 \omega_L y(0)  \left( 1-\frac{4\omega_L^2}{\Omega^2} \right).
\end{equation}
The direction of the drift depends on the sign of $y(0)$. Since the Hamiltonian is a quadratic polynomial in position and momentum, the classical velocity coincides with the motion of the center of a wave packet, which is launched at position $\mbfr(0)$ \cite{Ehrenfest1927}. However, the classical velocity is not identical to the quantum mechanical group velocity relevant for electronic transport.
The classical trajectories display a surprising variety of different behaviours: the number of orbits which return after some time $t$ to the launch point can be either finite (for $\frac{1}{2}m {y(0)}^2 \omega_y^2<E<\frac{1}{2}m {y(0)}^2 \Omega^2$) or infinite for $E>\frac{1}{2}m {y(0)}^2 \Omega^2$.

\subsection{The local density of states.}\label{sec:LDOSQM}

{\color{black}The quantum mechanical local density of states at point $\mathbf{r}$ is closely related to the classical orbits which lead from $\mathbf{r}$ to $\mathbf{r}$ (``closed orbits'', see Refs.~\cite{Berry1972,Peters1993}).
In quantum mechanics, the propagator (or Feynman kernel) $K$ advances the initial state $\psi(\mathbf{r},t')$ from initial time $t'$ to a later time $t$:
\begin{equation}
\psi(\mathbf{r},t)=\int_{-\infty}^\infty\rmd\mathbf{r}'\;K(\mathbf{r},t|\mathbf{r}',t')\psi(\mathbf{r}',t').
\end{equation}
The local density of states follows then from the relation 
(see Sect.~3.3.1 in Ref.~\cite{Kramer2005})
\begin{equation}
\label{eq:LDOS}
n(\mbfr;E) = \frac{1}{2\pi\hbar} 
\lim_{\eta\rightarrow 0}
\int_{-\infty}^\infty \rmd t\, \rme^{\rmi [E+\rmi\eta]t/\hbar}\, 
K(\mbfr, t|\mbfr,0).
\end{equation}
The propagator can be evaluated by summing the classical action $S_{\rm cl}$ over all possible paths from $\mbfr'$ to $\mbfr$
\begin{equation}
K(\mbfr, t|\mbfr',0)=1/N \int {\mathcal{D}\mbfr}\; \rme^{\rmi S_{\rm cl}(\mbfr, t|\mbfr',0)},
\end{equation}
where $N$ denotes a normalization factor and ${\mathcal{D}\mbfr}$ the path summation (see \cite{Grosche2013} for details and comprehensive compilation of analytically known Feynman path integrals).
}
In the absence of disorder and interactions, and without any external fields, the density of states in two-dimensions is independent of the energy, whereas a purely magnetic field reshapes the DOS to a series of discrete energy levels at energies $\hbar\omega_L(2n+1)$ \cite{Berry1981}. 
The continuous or discrete nature of the energy-spectrum reflects the fact that in the first case an electron with energy $E>0$ will travel arbitrary far from its initial position, whereas in a purely magnetic field its maximal distance is classically restricted to twice the cyclotron radius.

In the presence of a homogeneous electric field orthogonal to the magnetic field, a classical electron undergoes a uniform drift motion with a velocity given by the ratio of the electric and magnetic fields ${\cal E}/{\cal B}$. The corresponding LDOS becomes a smooth function of the energy \cite{Kramer2004}. Interestingly, between two former Landau levels the LDOS is extremely suppressed and thus a transport is not possible for energies in that range. This is in sharp contrast to the classical result, in which electrons propagate independent of their initial energy. Another suppression of the LDOS happens inside a former Landau level, which leads to the possibility of fractionally filled levels. The regions of strong transport blocking are transformed into plateaus of integer multiples of the conductivity quantum $e^2/h$ \cite{Kramer2005c}. In a homogeneous electric field, the degeneracy of a purely magnetic Landau level is lifted but the density of electrons per Landau level remains $N_{\cal B}=e {\cal B}/h$.
In the presence of a long range electrical potential, the \textit{local} density remains well defined, while one has to proceed carefully introduce the global density of states (see Ref.~\cite{Davies1998}, Sec.~6.2.1, and Ref.~\cite{Kramer2005c}, App.~C).

\subsubsection{Quantum mechanical expression for the LDOS.} 

\begin{figure}[t]
\begin{center}
\includegraphics[width=0.525\textwidth]{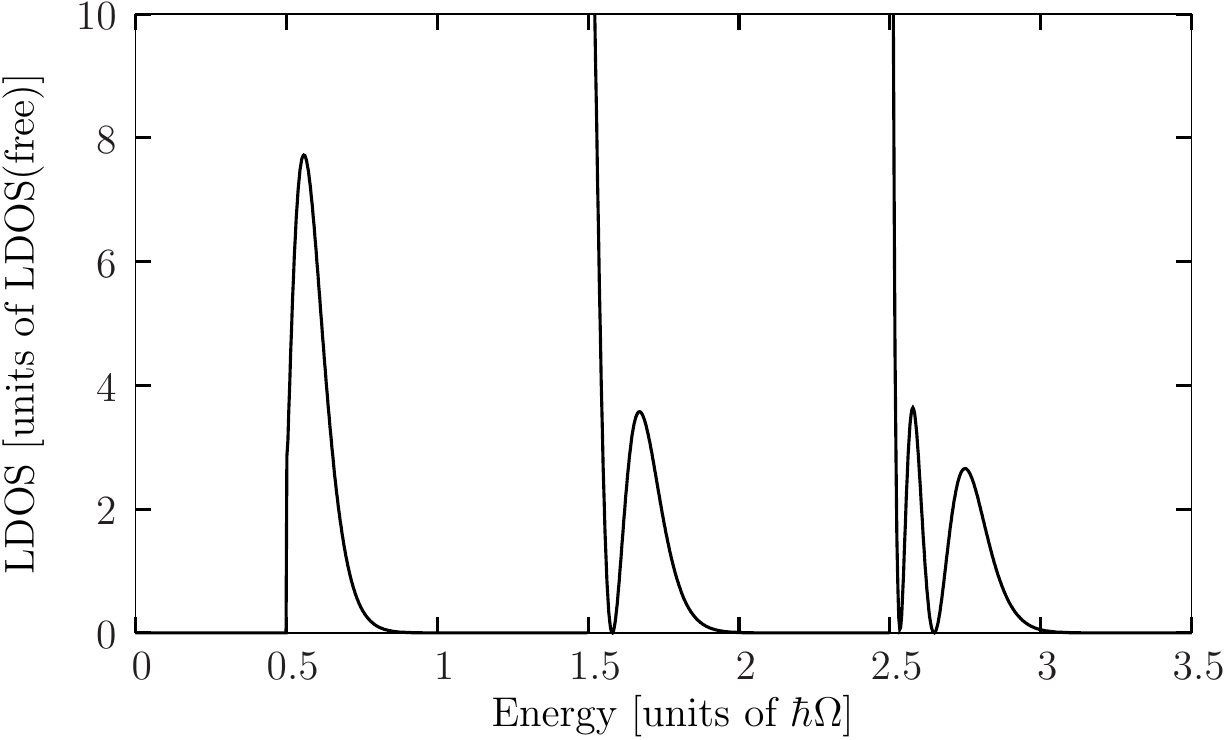}
\includegraphics[width=0.465\textwidth]{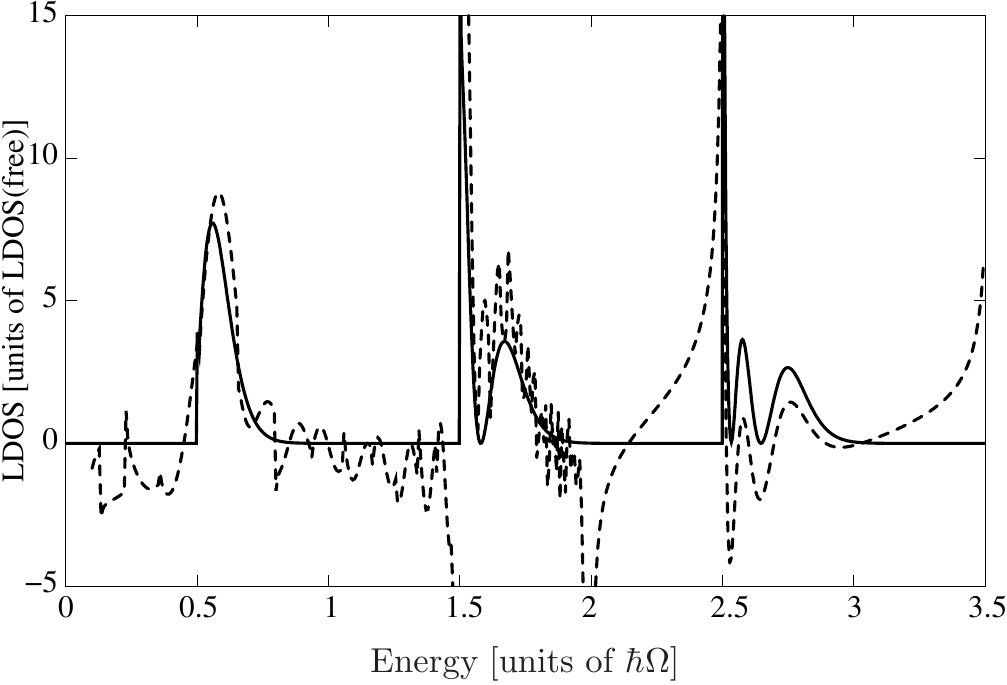}
\end{center}
\caption{
LDOS for a parabolic waveguide in a magnetic field. Parameters: ${\cal B}=5$~T, ${\cal E}=4000$~V/m, $A=0.4$, $\Omega=\omega_L\sqrt{4+A^2}\approx 2.04\;\omega_L$, $x=0, y=\frac{e{\cal E}}{m A^2 \omega_L^2}\approx 22.7$~nm. The LDOS is separated into distorted ``Landau levels'' which are again subdivided. Singularities at $\hbar\Omega(n+1/2)$ mark the onset of each Landau level.
Right panel: Comparison of the semiclassical LDOS (dashed line) and the quantum mechanical LDOS (solid line). Around $E=2\hbar\Omega$, the classical dynamics switches from a finite number of closed orbits to an infinite number.}
\label{fig:ldos}
\end{figure}
We briefly review the result for the local density of states, see also \cite{Kramer2010b} for a detailed derivation.
In the Landau gauge, the normalized eigenfunctions and eigenenergies of the Hamiltonian~(\ref{eq:HamiltonParab}) are:
\begin{equation}
\phi_{n,k_x}(x,y)=\frac{\rme^{\rmi k_x x}}{\sqrt{2\pi}}\frac{1}{\sqrt{2^n n! \sqrt{\pi}l}}
\exp\left[-\frac{{(y-\frac{2\omega_L}{\Omega^2}\frac{\hbar k_x}{m})}^2}{2 l^2}\right]
\HerH_n\left(\frac{y-\frac{2\omega_L}{\Omega^2}\frac{\hbar k_x}{m}}{l}\right),
\quad l=\sqrt{\frac{\hbar}{m\Omega}},
\end{equation}
\begin{equation}
E_{n,k_x}=\left(n+\frac{1}{2}\right)\hbar\Omega+\frac{\hbar^2 k_x^2}{2m}\frac{\Omega^2-4\omega_L^2}{\Omega^2}
\end{equation}
We express the LDOS by a sum and integral over eigenfunctions weighted with eigenenergies:
\begin{eqnarray}
n_{\rm parab}(\mbfr;E)
&=&\sum_{n=0}^{\infty}\int_{-\infty}^{\infty}\rmd k_x\,
\delta(E-E_{n,k_x}) {|\phi_{n,k_x}(x,y)|}^2\\
&=&
\sum_{n=0}^{\infty}\sum_{k=k_-,k_+} 
{\left|\frac{\partial E_{n,k_x}}{\partial k_x}\right|}^{-1}_{k_x=k}
\Theta(E-\hbar\Omega(n+1/2)) {|\phi_{n,k}(x,y)|}^2,\\
&&k_{\pm}=\pm\frac{\sqrt{2m [E-\hbar\Omega(n+1/2)]}\Omega}{\hbar\sqrt{\Omega^2-4\omega_L^2}}.
\end{eqnarray}
For each quantum number $n$, the energy integrated LDOS becomes
\begin{equation}
N_n=\int_{-\infty}^\infty\rmd E\;n_{{\rm parab},n}(\mbfr;E)
=\frac{\Omega^2}{4\omega_L^2} \frac{e {\cal B} }{2\pi\hbar}.
\end{equation}
Note that there is no spatial dependence of $N_n$ remaining after the energy integration. The density of electrons per quantized level differs from the quantization in a purely magnetic field (or orthogonal electric and magnetic fields) by the factor $\Omega^2/(4\omega_L^2)$. Also the LDOS has singularities at energies $E=\hbar\Omega(n+1/2)$. Typical graphs are shown in Fig.~\ref{fig:ldos}. 
The singularities in the LDOS are related to the localization (or ``trapping'') of part of a wavepacket, seen also in Fig.~\ref{fig:wp}, right panel: while the center of the wavepacket is drifting away, it expands also against the drift direction and always parts of the density remain at the origin.

\subsubsection{Semiclassical determination of the LDOS.}\label{sec:LDOSSC}

The semiclassical analysis of the problem provides a direct link between the classical electron drift trajectories and the quantum mechanical LDOS. 
Despite the possible presence of infinitely many orbits, it is possible to sum up their contributions and obtain the semiclassical result. 
Classical orbits, which lead from $\mbfr$ back to $\mbfr$, are best found by using the classical action. 
Each classical trajectory must be a extremal point of the reduced classical action $S_{\rm cl}$:
\begin{equation}\label{eq:SaddlePoints}
\frac{\partial S_{\rm cl}(\mbfr,t|\mbfr,0)}{\partial t}+E=0.
\end{equation}
For the Hamiltonian (\ref{eq:HamiltonParab}), the classical action is given by
\begin{equation}\label{eq:SclClosed}
S_{\rm cl}(\mbfr, t|\mbfr,0)=-\frac{m y^2 \Omega^2}{8\omega_L^2/[(\Omega^2-4\omega_L^2)t]+\Omega\cot(\Omega t/2)}.
\end{equation}
The long-time behaviour is readily extracted by setting the first term in the denominator to zero
\begin{equation}\label{eq:SclClosedAsym}
S_{\rm cl}^{\rm as}(\mbfr, t|\mbfr,0)=-m y^2 \Omega\tan(\Omega t/2).
\end{equation}
In the asymptotic limit, the saddle points are available analytically by inserting (\ref{eq:SclClosedAsym}) into  (\ref{eq:SaddlePoints}):
\begin{eqnarray}
t_n^\cup  &=&\frac{2n\pi-\arccos(m y^2 \Omega^2/E-1)}{\Omega}, \quad n=1,2,3,\ldots\\
t_n^\cap&=&\frac{2n\pi+\arccos(m y^2 \Omega^2/E-1)}{\Omega}, \quad n=0,1,2,\ldots.
\end{eqnarray}
Here $(t_n^\cup,t_n^\cap)$ denotes a minimum and maximum of the classical action functional. Note that in each interval $[t_n,t_{n+1}]$ a singularity exists. 
%In Fig.~\ref{fig:compare} we compare the asymptotic value of the classical action with the exact expression. 
For $\frac{1}{2} m y^2\Omega^2-E>0$, the long time asymptotic fails and after a certain time no classical trajectories prevail. The number of closed orbits is given by a finite number.

Next we express the local density of states in terms of the propagator
\begin{equation}
\label{eq:DOSIntegralParab}
n_{\mgf,y^2}^{(2D)}(\mbfr;E) = \frac{1}{2\pi\hbar} 
\lim_{\eta\rightarrow 0}
\int_{-\infty}^\infty \rmd t\, \rme^{\rmi [E+\rmi\eta]t/\hbar}\, 
K_{\mgf,y^2}(\mbfr, t|\mbfr,0),
\end{equation}
where $K_{\mgf,y^2}$ is given in closed form by
\begin{eqnarray}\label{eq:Kparab}
K_{\mgf,y^2}(\mbfr, t|\mbfr,0) &=& a(t) \exp \left\{ \frac\rmi\hbar S_{{\rm cl}}(\mbfr, t|\mbfr,0)\right\},\\
a(t)&=&\frac{m\Omega^2 {(2\pi\rmi\hbar)}^{-1} \sgn(t_{{\rm sing},[\Omega t/(4\pi)]+1}-t)}
{\sqrt{2\sin(\Omega t/2)  }\sqrt{\Omega t(\Omega^2-4\omega_L^2)\cos(\Omega t/2)+8\omega_L^2\sin(\Omega t/2)}}\nonumber.
\end{eqnarray}
Note the phase correction in the propagator $\sgn(t_{{\rm sing},[\Omega t/(4\pi)]+1}-t)$,
which arises from the use of the square root in the propagator. The symbol $[\cdots]$
denotes the integer part of a number and is used to generate an index to the list of zeroes
of the expression
\begin{equation}
\Omega t(\Omega^2-4\omega_L^2)\cos(\Omega t/2)+8\omega_L^2\sin(\Omega t/2)=0,
\end{equation}
where the zeroes are numbered such that $t_{{\rm sing},n}$ is contained in the interval
$(\frac{4\pi}{\Omega}(n-1/2),\frac{4\pi}{\Omega}n)$. Another set of singularities occurs at
$t=\frac{2\pi n}{\Omega}$, $n\in\mathbf{N}$.
\begin{figure}[t]
\begin{center}
\includegraphics[width=0.7\textwidth]{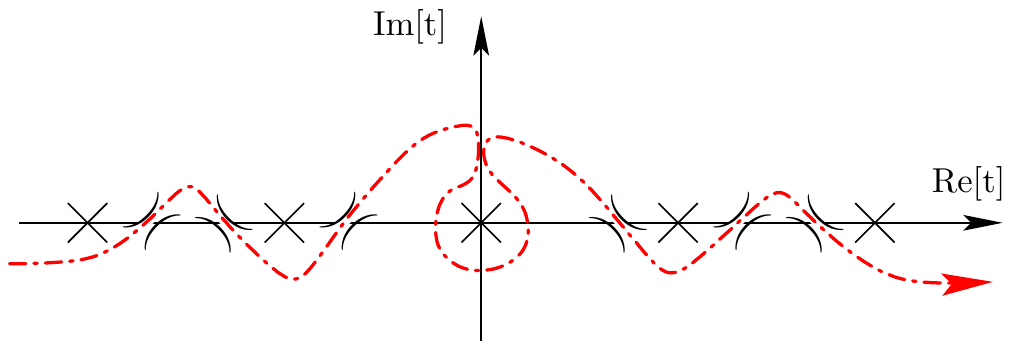}
\end{center}
\caption{
Principal structure of the classical action in the complex time-plane. The dashed line denotes the integration path. Singularities are denoted by $\times$ and saddle points by $)($. Note that the singularity at the origin arises from the prefactor $a(t)$ in the propagator (\ref{eq:Kparab}) and not from the classical action at $t=0$.
\label{fig:integrationpath}}
\end{figure}

An asymptotic evaluation of the integral~(\ref{eq:DOSIntegralParab}) provides the link between closed orbits and the density of states. The original path of integration follows the real time-axis. An analytic continuation of the propagator makes it possible to deform this path of integration to the one sketched in Fig.~\ref{fig:integrationpath}. The new integration contour passes through saddle points of the exponent [denoted by $)($ in the figure] using the paths of steepest descent. The singularities of the integrand are denoted by $\times$ in Fig.~\ref{fig:integrationpath}. The unavoidable contribution of the singularity at $t=0$, may be evaluated by the residue-theorem:
\begin{equation}
I_{O}
=\frac{1}{2\pi\hbar}\oint\rmd t\, 
\rme^{\rmi E t/\hbar}\, K_{\mgf,y^2}(\mbfr, t|\mbfr,0)
=\frac{m}{2\pi\hbar^2}.
\end{equation}
This contribution due to the short time-limit $t\rightarrow 0$ of the propagator equals the LDOS in the absence of any external potentials.
Combining its value with the contributions from the saddle-points at time $t_k$ yields the semiclassical result:
\begin{equation}
n_{sc,\mgf,y^2}^{(2D)}(\mbfr;E)=I_{O}+2\,{\rm Re}\left[\frac{1}{2\pi\hbar}\sum_{k=1}^\infty
a(t_k)\frac{\rme^{\rmi E t_k/\hbar+\rmi S_{cl}(\mbfr,t_k|\mbfr,0)/\hbar+\rmi\pi{\rm  
sgn}[\ddot{S}_{cl}(\mbfr,t_k|\mbfr,0)]/4}}{\sqrt{|\ddot{S}_{cl}(\mbfr,t_k|\mbfr,0)|/(2\pi\hbar)}}
\right]
\end{equation}
The main difficulty comes from the infinite number of orbits, which have to be summed up. Noting that there exists an analytic expression for the contribution of each closed orbit, we can actually perform the summation.
Asymptotically, the values of the classical action of the $n$th pair are
\begin{equation}
S_{\rm cl}^{{\rm as},\cup,\cap}(E)=\pm y \sqrt{m(2E-my^2\Omega^2)}
\end{equation}
The sum over closed orbits can be expressed with the Lerch zeta function $\LerchL(\lambda,\alpha,s)$, or the Lerch transcendent $\Phi(z,s,\alpha)$ \cite{Bateman1953}:
\begin{equation}
\LerchL(\lambda,\alpha,s)=\sum_{n=0}^{\infty}\frac{\exp(2\pi\rmi\lambda n)}{{(n+\alpha)}^s},
\quad
\Phi(z,s,\alpha)=\sum_{n=0}^{\infty}\frac{z^n}{{(n+\alpha)}^s},
\quad
\Phi(\rme^{2\pi\rmi\lambda},s,\alpha)=\LerchL(\lambda,\alpha,s).
\end{equation}
The contribution of the orbits which lead to a local minimum in the classical action is given by
\begin{eqnarray}
n_\infty^{{\rm as},\cup}(y;E)&=&
2\,{\rm Re}
\left[\frac{\exp[\rmi S_{\rm cl}^{{\rm as},\cup}(E)/\hbar+\rmi\pi/4]}{\sqrt{2\pi\hbar}}
\sum_{k=1}^\infty \frac{a(t^\cup) \exp[\rmi E t^\cup/\hbar]}
{\sqrt{|\ddot{S}_{cl}(\mbfr,t_k|\mbfr,0)|}}
\right]\nonumber\\
&=&
2\,{\rm Re}\bigg[\frac{\exp[\rmi S_{\rm cl}^{{\rm as},\cup}(E)/\hbar
-\rmi E \arccos(my^2\Omega^2/E-1)/(\hbar\Omega)
+\rmi\pi/4]m\Omega^{3/2}}{\sqrt{2\pi\hbar}\hbar\sqrt{(2E-my^2\Omega^2)(\Omega^2-4\omega_L^2)}\sqrt{2\pi}}\nonumber\\
&&\times
\sum_{k=1}^\infty \frac{{(-1)}^k\exp[2\pi\rmi k E/(\hbar\Omega)]}{\sqrt{k+\arccos(my^2\Omega^2/E-1)/(2\pi)}}
\bigg]\\\nonumber
&=&b(y,E)\;\left[
\LerchL\left(\frac{E}{\hbar\Omega}+\frac{1}{2},\frac{\arccos(my^2\Omega^2/E-1)}{2\pi},\frac{1}{2}\right)
-\frac{\sqrt{2\pi}}{\sqrt{\arccos(my^2\Omega^2/E-1)}}\right]
\end{eqnarray}
The Lerch zeta function is defined as the limit
\begin{equation}
\lim_{\eta\rightarrow 0}
\LerchL\left(\frac{E+\rmi\eta}{\hbar\Omega}+\frac{1}{2},\frac{\arccos(my^2\Omega^2/E-1)}{2\pi},\frac{1}{2}\right)
\end{equation}
A similar expression holds for the contributions of the local maxima. The semiclassical LDOS has a divergence at energies $E=(n+1/2)\hbar\Omega$, which exactly corresponds to the quantum mechanical result.
A comparison of the semiclassical and the quantum mechanical result is shown in the right panel of Fig.~\ref{fig:ldos}.
{\color{black}The semiclassical LDOS correctly identifies the onset of each Landau level and the substructure within the level, but additionally introduces fast oscillations due to the changing number of extremal paths.
A standard remedy for the resulting divergences at caustics is to employ a uniform approximation \cite{Bracher2006}, which smoothens the result.
The prominent divergence of the semiclassical result at $E=2\hbar\Omega$ originates from the transition of a finite number of closed orbits to an infinite number of terms.
The evaluation of the semiclassical LDOS shows that already the extremal paths of the classical action are in general sufficient to  reproduce the mixed continuous and discrete spectrum of the electron confined to a wave-guide in a magnetic field.
In experiments, the discrete values signal the opening of a new quantized channel and lead to a stepwise conductivity seen in nanostructures \cite{Kramer2010}.
Next, we discuss how a similar transition between trapped and escaping states arises in celestial mechanics.
}

\section{Comets orbiting Jupiter: the three-body problem.}\label{sec:3body}

\begin{figure}[bt]
\begin{center}
\includegraphics[height=0.292\textwidth]{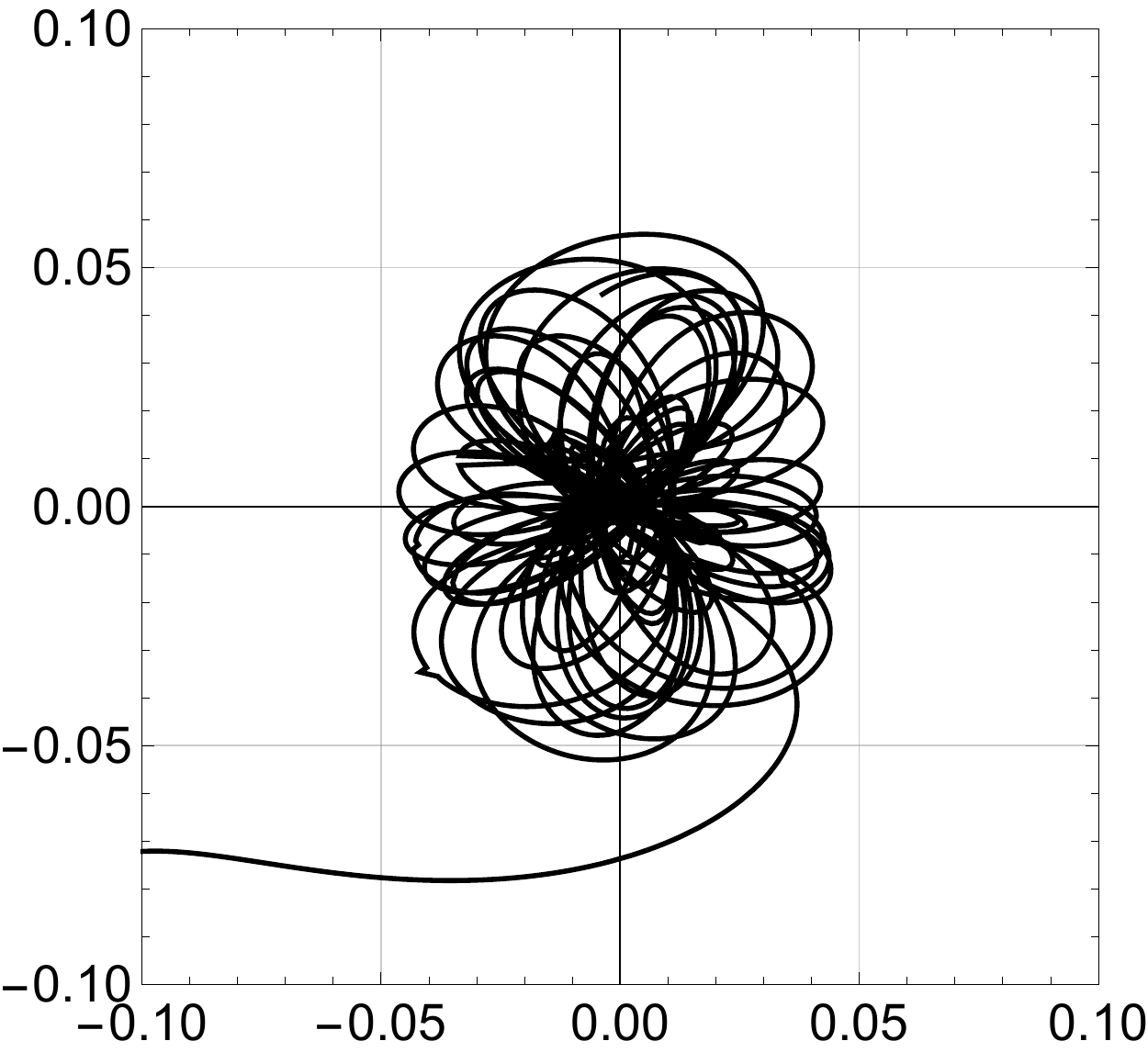}
\includegraphics[height=0.292\textwidth]{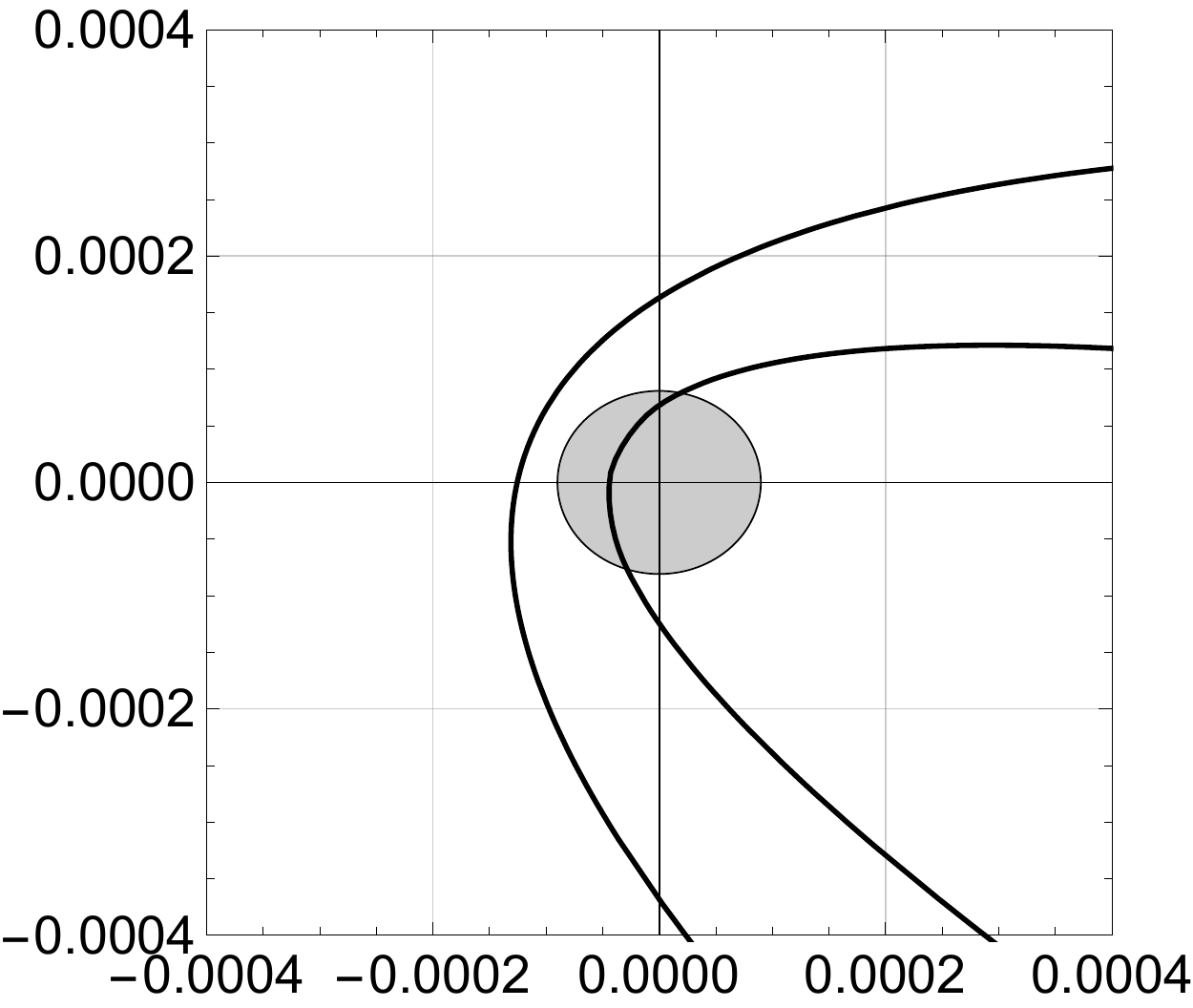}
\includegraphics[height=0.292\textwidth]{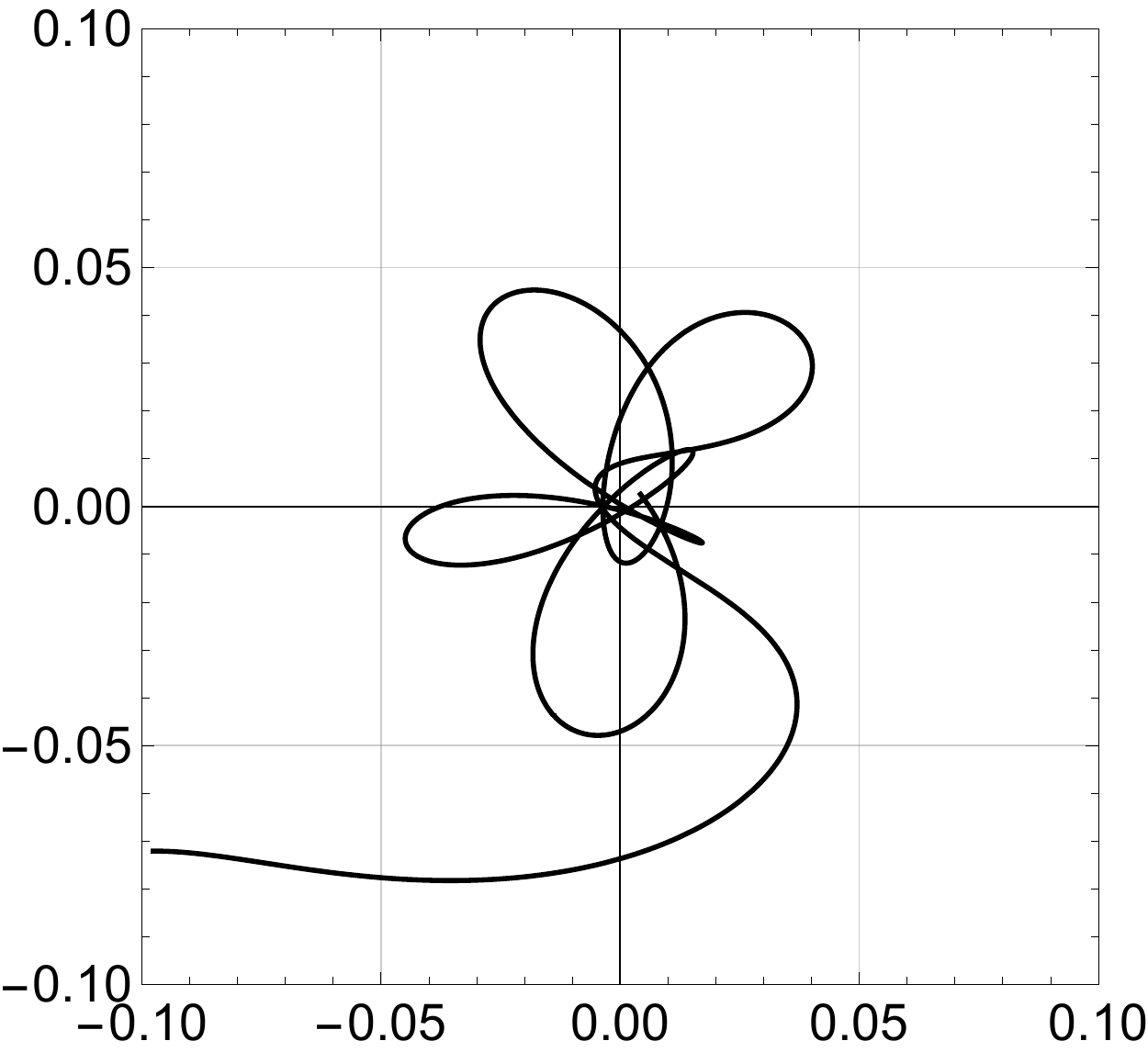}
\end{center}
\caption{
Numerical integration of comet SL9 K fragment (JPL Solar System Dynamics Group SPICE kernel for Shoemaker-Levy 9 (D/1993 F2-K)).
Left panel orbit ($x-y$ plane) from the year 1900 until 2014, middle panel:
close approach 1992 and impact in July 1994 (Jupiter disk marked, $x-z$ plane), the numerical integrations sees Jupiter as a point mass and continues in time.
Right panel: escape of SL9 from Jupiter 2014 ($x-y$ plane). A time-reversed version would show a possible capture scenario for SL9.
All units: fractions of Jupiters semi-major axis, Jovicentric system, left and right panels: $y$-axis pointing away from the sun, $z$-axis normal to Jupiter orbital plane. 
\label{fig:sl9spice}}
\end{figure}

\begin{figure}[bt]
\begin{center}
\includegraphics[width=0.6\textwidth]{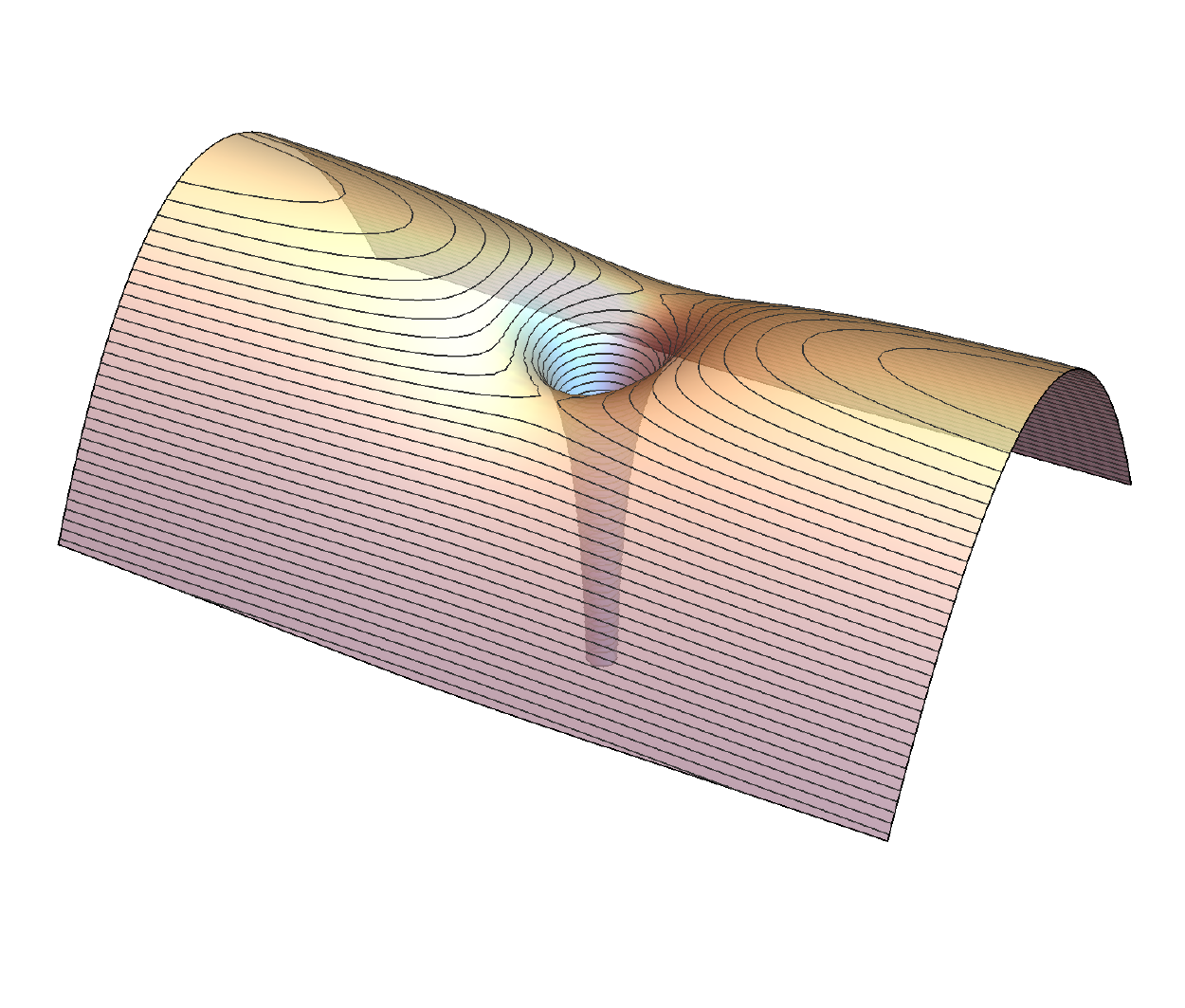}
\end{center}
\caption{
Visualization of the potential energy surface $V_{\rm pot}=-\frac{\mu_j}{\sqrt{x^2+y^2}}-\frac{\omega_y^2 y^2}{2}$ for $\omega_y^2=3$, $\mu_j=1/1000$.
\label{fig:potsaddle}}
\end{figure}

A modification of the previous problem leads to a form 
applicable to the restricted three-body problem in celestial mechanics.
{\color{black}
We change the sign of the parabolic confinement potential and add an additional attractive Coulomb potential (strength $\kappa$).
In view of the previous section, the added Coulomb potential could be seen as an impurity present in the wave guide, see Fig.~\ref{fig:potsaddle}.
The magnetic field part is unchanged, but for a more convenient connection to celestial mechanics now written in the symmetric gauge (vector potential $\AVP=\frac{1}{2}\MGF\times\mathbf{r}$, magnetic field $\MGF=\mgf \hat{\mathbf{z}}$).
}
The modified Hamiltonian reads 
\begin{equation}\label{eq:HamCom}
H=\frac{1}{2m}(p_x^2+p_y^2)
+\frac{1}{2}m \omega_l^2 (x^2+y^2)
+\omega_l (p_x y - p_y x)
-\frac{1}{2}m \omega_y^2 y^2
-\frac{\kappa}{\sqrt{x^2+y^2}}.
\end{equation}
The classical equations of motions then become
\begin{eqnarray}
\ddot{x}&=&-\kappa \frac{x}{{(x^2+y^2)}^{3/2}}+2\omega_l \dot{y}\label{eq:ceq1}\\
\ddot{y}&=&-\kappa \frac{y}{{(x^2+y^2)}^{3/2}}-2\omega_l \dot{x}-\omega_y^2 y.\label{eq:ceq2}
\end{eqnarray}
Closely related equations appear in the planar three-body problem of a small body encircling a planet, which is in turn orbiting the sun (see for instance the extensive discussion and formulae provided in the JPL Technical report \cite{Broucke1968}).
{\color{black}In the celestial mechanics of planets, magnetic fields are usually not considered for the orbital integration.
However, similar velocity dependent terms to Eqs.~(\ref{eq:ceq1},\ref{eq:ceq2}) arise from the transformation of the classical equations of motion of planets to a rotating frame.}
We adjust the period of the rotating frame to the angular frequency of Jupiter around the barycenter of the Jupiter-Sun system (synodal reference frame).
In recent literature (see \cite{Anderson2019}, Eqs.~(1,2)) this system is denoted by 
\begin{eqnarray}
\ddot{x}&=&\partial_x \phi(x,y)-2 \dot{y}\label{eq:eomHamCom1}\\
\ddot{y}&=&\partial_y \phi(x,y)+2 \dot{x}\label{eq:eomHamCom2},
\end{eqnarray}
with Jupiter of mass $\mu_j$ located at $x_j=1-\mu_j$, $y_j=0$, and the sun with mass $1-\mu_j$ at $x_s=-\mu_j$, $y_s=0$.
This leads to a total mass of $1$ and the angular velocity of the rotating frame becomes $1$.
The potential is given by
\begin{equation}
\phi(x,y)=\frac{1}{2}(x^2+y^2)+\frac{\mu_s}{\sqrt{{(x-x_s)}^2+{(y-y_s)}^2}}+\frac{\mu_j}{\sqrt{{(x-x_j)}^2+{(y-y_j)}^2}},
\end{equation}
{\color{black}where the quadratic terms account for the centrifugal contributions within the rotating frame.
Additionally we shift the origin of the coordinate system to the center of Jupiter $(x,y)\rightarrow(x+x_j,y+y_j)$ and obtain the equation of motions:
\begin{eqnarray}
\ddot{x}&=&(x+x_j)
-\frac{x \mu_j}{{\left(x^2+y^2\right)}^{3/2}}
-\frac{(x-x_j+x_s) \mu_s}{{\left(x^2+y^2\right)}^{3/2}}-2\dot{y}\\
\ddot{y}&=&(y+y_j)
-\frac{y \mu_j}{{\left(x^2+y^2\right)}^{3/2}}
-\frac{(y-y_j+y_s) \mu_s}{{\left(x^2+y^2\right)}^{3/2}}+2\dot{x}.
\end{eqnarray}
Upon setting expanding the solar acceleration terms with $\mu_s$ around Jupiter $(x,y)=(0,0)$ to first order and using $x_j-x_s=1, y_j=0$ we obtain the following equations of motions
\begin{eqnarray}
\ddot{x}&=&(1+2\mu_s)x-\frac{\mu_j x}{{(x^2+y^2)}^{3/2}}-2 \dot{y}\\
\ddot{y}&=&(1-1\mu_s)y-\frac{\mu_j y}{{(x^2+y^2)}^{3/2}}+2 \dot{x}.
\end{eqnarray}
By letting $m=1$, $q=1$, $\mgf=2$, $\omega_y=(1+2\mu_s)$, $\kappa=\mu_j$, and $\mu_j=1/1000 \ll 1$, $\mu_s=999/1000\approx 1$ in Eqs.~(\ref{eq:eomHamCom1},\ref{eq:eomHamCom2}) we find the mapping of the restricted three-body problem to the equivalent system of an electron in an attractive Coulomb potential and a perpendicular homogeneous field plus parabolic confinement.
}

\newpage

\subsection{Chaotic motion and transient trapping}

The solutions of equations similar to Eq.~(\ref{eq:HamCom}) have received considerable attention in celestial and quantum mechanics, respectively.
In quantum mechanics, the electronic states of hydrogen in the presence of a magnetic field are a prime example of the implications of classical chaotic motion on the eigenstates of a quantum mechanical problem \cite{Friedrich1989}.
{\color{black}Here, the presence of a parabolic waveguide puts the dynamics closer to the setups experimentally studied in semiconductor hetero\-structures.
An experimental realization is provided by quantum dots embedded in the arms of an interferometer \cite{Kreisbeck2017}, or an added impurity.}

\begin{figure}[bt]
\begin{center}
\includegraphics[width=0.45\textwidth]{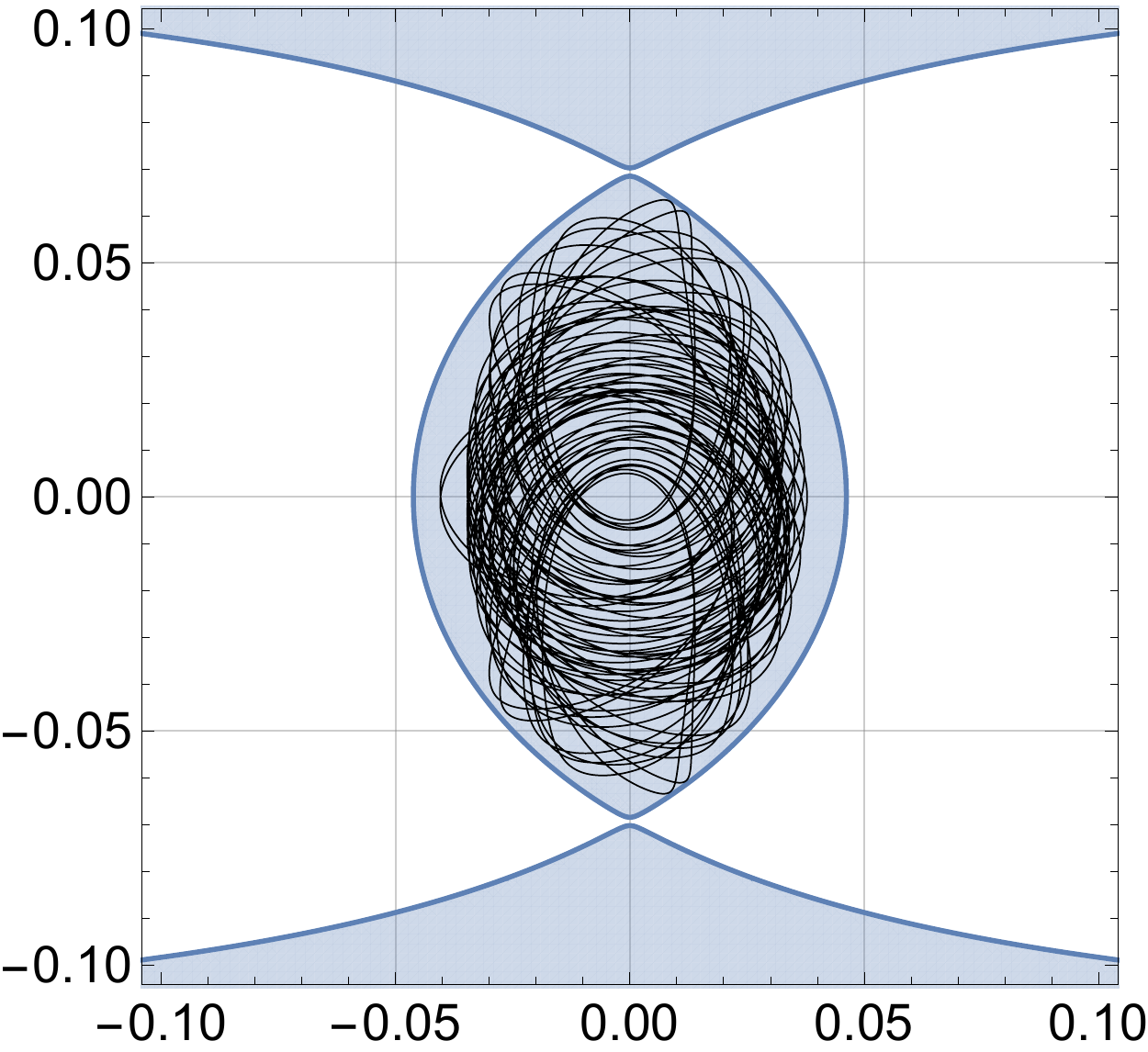}\hfill
\includegraphics[width=0.45\textwidth]{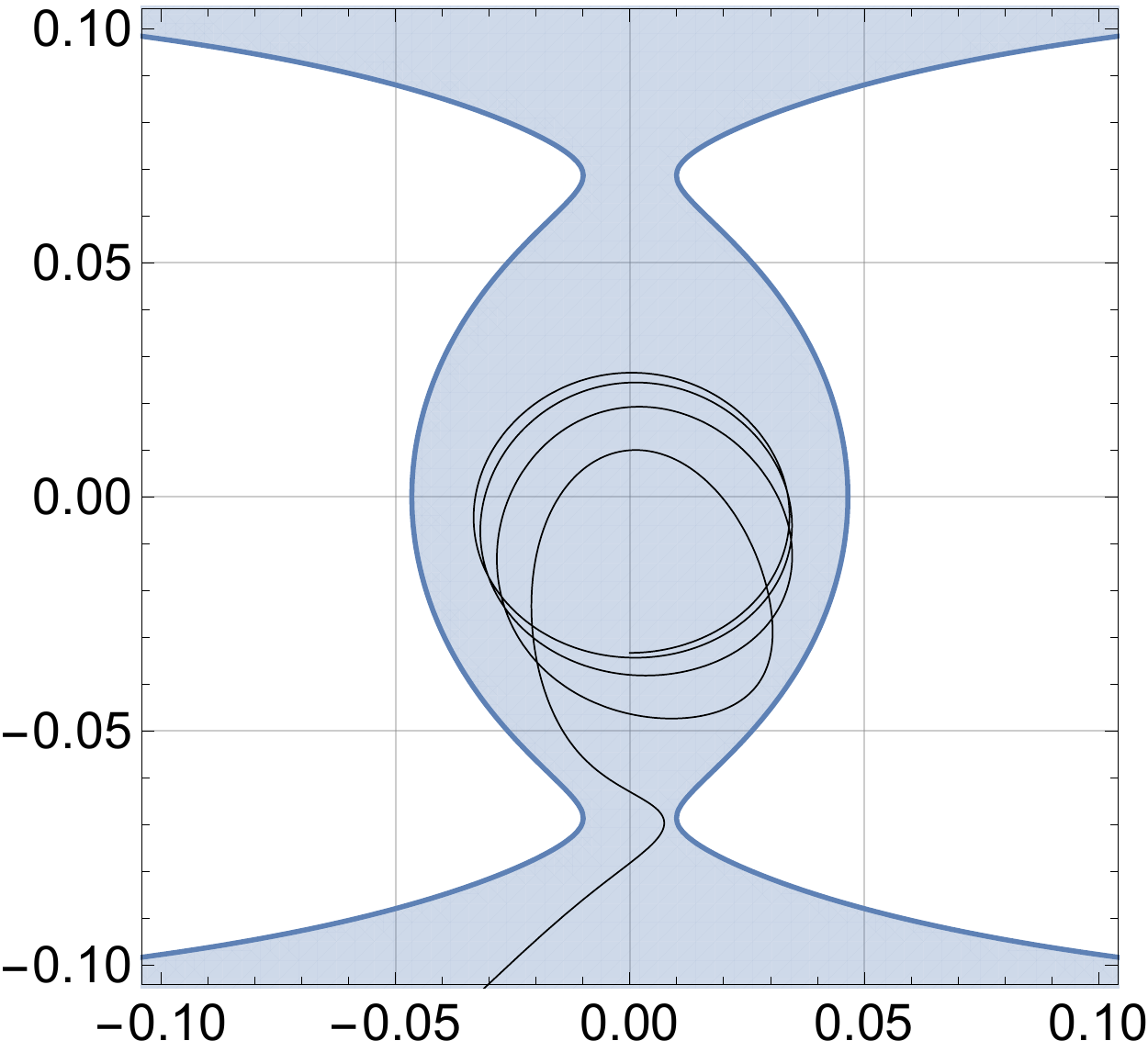}
\end{center}
\caption{
Classical trajectory $(x_0,y_0)=(0,-1/30)$, $(\dot{x}_0,\dot{y}_0)=(k_0,0)$ for the Hamiltonian (\ref{eq:HamCom}).
Left panel: trapped orbit $k_0=14163/100000$, right panel: escape after many times encircling Jupiter ($k_0=1427/10000$).
The shaded area shows the regions which are energetically accessible.
\label{fig:escape}}
\end{figure}

{\color{black}The prototypical application of  Eq.~(\ref{eq:HamCom}) in the context of celestial mechanics is the transient capturing of comets around a larger parent body, mostly Jupiter, but also the Earth (asteroid $2020 CD_3$).
The restriction to a geometry with all three bodies (sun, planet, comet) in one plane allows us to gain easier physical insight into the intricate dynamics.
}
Comet Shoemaker-Levy 9 (SL9) is one of the best known examples of a comet encircling Jupiter for many decades before crashing into the atmosphere \cite{Benner1995}. 
Comet SL9 disintegrated into several pieces due to tidal forces at an earlier close approach to Jupiter in 1992 (see Fig.~\ref{fig:sl9spice}) and a whole string of particles left a chain of impact locations on the fast spinning Jupiter in July 1994, witnessed by astronomers.

Under the assumption of a point-sized Jupiter, SL9 would have continued its orbit after July 1994 and would have left the transient orbit around Jupiter in the year 2014. 
Fig.~\ref{fig:sl9spice} shows the orbital integration of the K-fragment of comet SL9.
To understand the transient trapping of an object in the vicinity of Jupiter it is helpful to investigate the Hamiltonian (\ref{eq:HamCom}) in more detail. 
This simplified setup shows a very similar dynamics and is highly sensitive to the initial conditions, see Fig.~\ref{fig:escape}.
The complex dynamics is caused by the vicinity of comet SL9 to the Lagrange $L_1$ and $L_2$ points, where the forces of Sun and Jupiter in a rotating frame with Jupiter cancel the centrifugal force.
Another force, which is neglected here, is the non-gravitational force arising from the sublimation of ices on the cometary nuclei and changes the cometary orbit as described by Bessel \cite{Bessel1836}.
A well-studied example is the comet 67P/Churyumov-Gerasimenko \cite{Kramer2017,Lauter2019}, the target object of the Rosetta mission.
There, the momentum transferred by the sublimating molecules shortened the $\approx 12$~h rotation period by 20~minutes, tilted the rotation axis, and changed the orbit at the 2015~apparition \cite{Kramer2019,Kramer2019a}.

\subsection{Quantum mechanical treatment of the orbit of comet SL9: spectrum and eigenfunctions}

To be clear: we do not claim that comet SL9 has to be treated as a quantum-mechanical wave-packet.
Rather we highlight the close connection of celestial mechanics to the equivalent quantum mechanical solution. 
This approach was pioneered by Eric~Heller \cite{Heller1978} and is reviewed in detail in \cite{Kramer2008,Kramer2011a}.
It is helpful to introduce the energy dependent Green's function
\begin{equation}
G(\mbfr,\mbfr';E)=
\sum_{n}\frac{\psi_n(\mbfr)\psi_n(\mbfr')}{E-E_n}
+\int\rmd\nu\frac{\psi_\nu(\mbfr)\psi_\nu(\mbfr')}{E-E_\nu}.
\end{equation}
This allows us to express the expectation value of the Green's function with respect to a spatially extended wavepacket $\psi(\mbfr)$ as 
\begin{eqnarray}\label{eq:ldoswp}
\langle \psi | G(E) | \psi \rangle
&=&\int \rmd\mbfr \int \rmd\mbfr' \psi(\mbfr,0)^* G(\mbfr,\mbfr';E) \psi(\mbfr',0)\\
&=&\frac{1}{\rmi\hbar}\int_0^\infty\rmd t\;\rme^{\rmi E t/\hbar}
\int \rmd\mbfr \int \rmd\mbfr' \psi(\mbfr,0)^*K(\mbfr,t|\mbfr',0) \psi(\mbfr',0) \\
&=&\frac{1}{\rmi\hbar}\int_0^\infty\rmd t\;\rme^{\rmi E t/\hbar}
\int \rmd\mbfr\; \psi(\mbfr,0)^* \psi(\mbfr,t) \\
&=&\frac{1}{\rmi \hbar}\int_0^\infty\rmd t\;\rme^{\rmi E t/\hbar} C(t),\label{eq:autocorr_laplace}
\end{eqnarray}
where $C(t)$ denotes the autocorrelation function
%
%\begin{equation}
$C(t)=\int \rmd\mbfr\; \psi(\mbfr,0)^* \psi(\mbfr,t)$,
%\end{equation}
and $K$ the time propagator. 

\begin{figure}[h!]
\begin{center}
\includegraphics[width=0.95\textwidth]{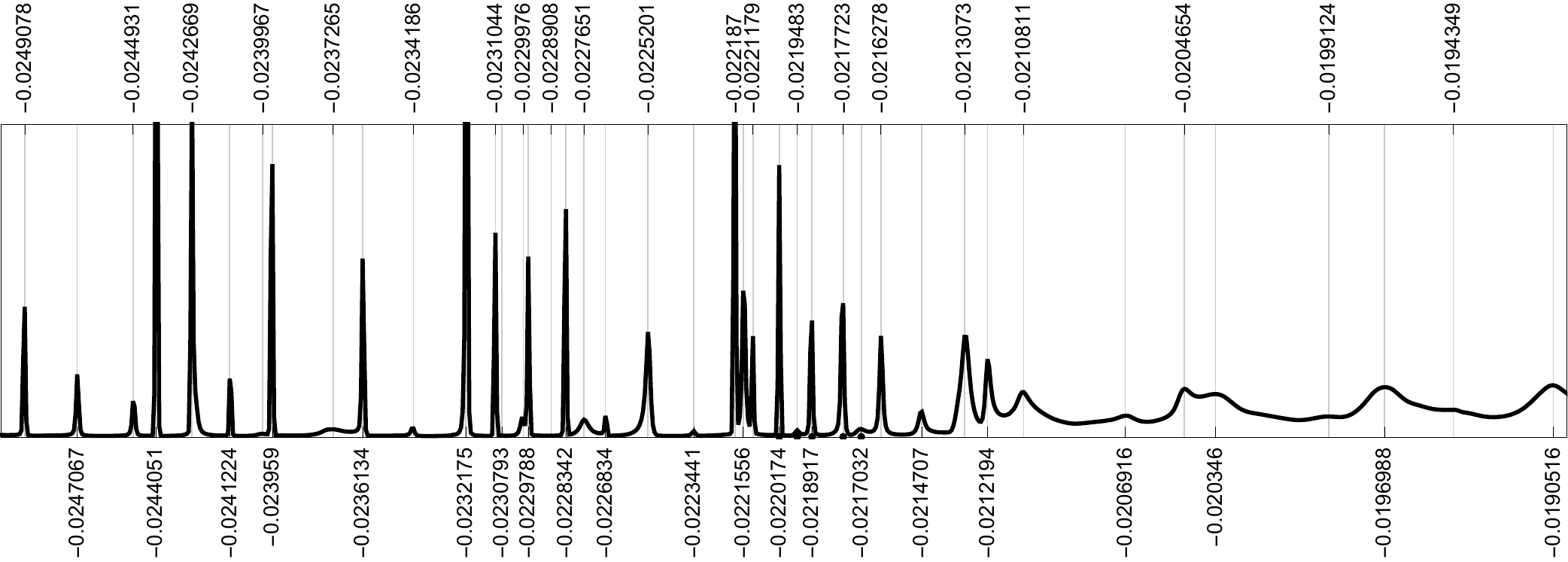}
\includegraphics[width=0.95\textwidth]{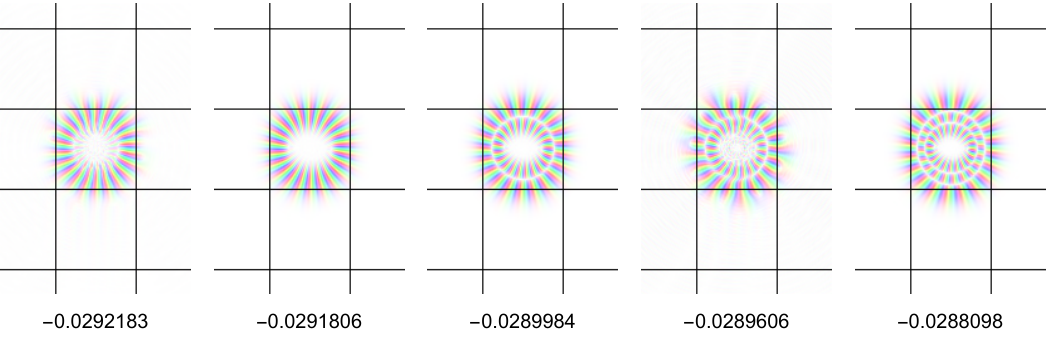}
\includegraphics[width=0.95\textwidth]{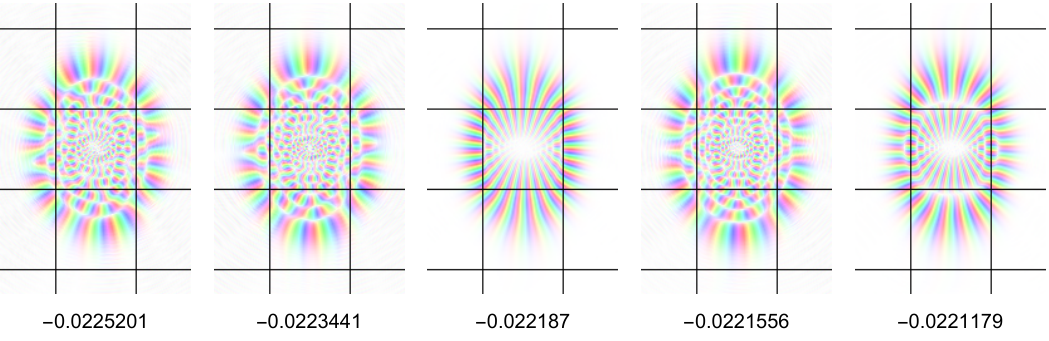}
\includegraphics[width=0.95\textwidth]{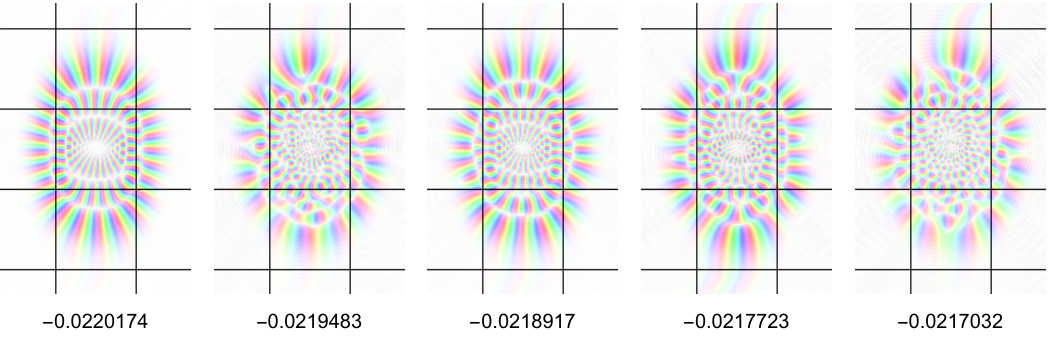}
\end{center}
\caption{
Upper panel: Fourier transform of the autocorrelation function with peaks identified by their energy ($\hbar=1/5000$) in the transition region between bound and continuum states.
Lower panel: energy states for 15 exemplary cases.
Parameters: $\hbar=1/5000$, grid cells $0.05 \times 0.05$, energies indicated.
\label{fig:spec}}
\end{figure}

\clearpage
\noindent
A convenient way to obtain the eigenfunctions of an open quantum system is to propagate a wave packet and to project out the eigenfunctions at peaks of the Fourier transform of the time-dependent autocorrelation function $C(t)$.
In the upper panel Fig.~\ref{fig:spec} we show the spectrum obtained from the propagation of a wavepacket  on a $512\times512$ lattice, centered initially at the same starting point as the classical trajectory.
The lower panels show the Green's function at the indicated energies.
The various eigenstates are clearly visible, including states where the electronic wave function leaks out from the nucleus (see for instance in the lower panels).
In contrast to classical dynamics, tunneling opens another pathway for the electronic charges to escape in the quantum case.

\section{Conclusion}

{\color{black}
We have explored the connection between cometary orbits around Jupiter and electron dynamics in a magnetic field.
Within the parameters considered (Taylor expansion of the solar potential, planar motion, orbit perpendicular to the angular momentum), both hamiltonians coincide.
In both systems, localized (bound) states emerge besides a continuous spectrum.
The transient trapping is augmented in quantum mechanics by classically not allowed tunneling.
For the hydrogen atom and also for molecular states the common analysis of classical and quantum dynamics has propelled the development of numerical and analytical methods.
Phase space studies are commonly conducted in different fields of physics, in particular in celestial dynamics.
In addition to providing physical insights into quantum states, they also benefit from the tools developed independently in quantum and semiclassical physics.
Semiclassical methods continue to provide a unified view of various fields of physics such as nano- and astrophysics \cite{Heller2019}.
}

\ack

I am indebted to Prof.~Eric~J.~Heller and for Prof.~Manfred~Kleber for sharing their insights into the (semi)classical--quantum connection and wave packet dynamics.
This work was supported over the years by the German Research Foundation (DFG) grant KR~2889.
I acknowledge compute time allocation by the North-German Supercomputing Alliance (HLRN).
Finally I thank the organizers of the ``Symmetry in Science'' symposium Bregenz~2019 (Prof.~Dieter Schuch and Prof.~Michael Ramek) for the invitation to present this work at the meeting.


\section*{References}


\begin{thebibliography}{10}
\expandafter\ifx\csname url\endcsname\relax
  \def\url#1{\texttt{#1}}\fi
\expandafter\ifx\csname urlprefix\endcsname\relax\def\urlprefix{URL }\fi
\expandafter\ifx\csname doiprefix\endcsname\relax\def\doiprefix{DOI }\fi
\providecommand{\bibinfo}[2]{#2}
\providecommand{\eprint}[2][]{\url{#2}}

\bibitem{Fabrikant1980}
\bibinfo{author}{Fabrikant, I.}
\newblock \bibinfo{title}{Interference effects in photodetachment and
  photoionization of atoms in a homogeneous electric field}.
\newblock \emph{\bibinfo{journal}{JETP (Russian original - ZhETF, Vol. 79, No.
  6, p. 2070, December 1980)}} \textbf{\bibinfo{volume}{52}},
  \bibinfo{pages}{1045} (\bibinfo{year}{1980}).

\bibitem{Demkov1982}
\bibinfo{author}{Demkov, Y.}, \bibinfo{author}{Kondratovich, V.} \&
  \bibinfo{author}{Ostrovskii, V.}
\newblock \bibinfo{title}{Interference of electrons resulting from the
  photoionization of an atom in an electric field.}
\newblock \emph{\bibinfo{journal}{JETP Lett}} \textbf{\bibinfo{volume}{34}},
  \bibinfo{pages}{403} (\bibinfo{year}{1982}).

\bibitem{Peters1993}
\bibinfo{author}{Peters, A.~D.} \& \bibinfo{author}{Delos, J.~B.}
\newblock \bibinfo{title}{Photodetachment cross section of {{H}} - in crossed
  electric and magnetic fields. {{I}}. {{Closed}}-orbit theory}.
\newblock \emph{\bibinfo{journal}{Physical Review A}}
  \textbf{\bibinfo{volume}{47}}, \bibinfo{pages}{3020--3035}
  (\bibinfo{year}{1993}).
\newblock \doiprefix 10.1103/PhysRevA.47.3020.

\bibitem{Bordas1998}
\bibinfo{author}{Bordas, C.}
\newblock \bibinfo{title}{Classical motion of a photoelectron interacting with
  its ionic core: {{Slow}} photoelectron imaging}.
\newblock \emph{\bibinfo{journal}{Physical Review A}}
  \textbf{\bibinfo{volume}{58}}, \bibinfo{pages}{400--410}
  (\bibinfo{year}{1998}).
\newblock \doiprefix 10.1103/PhysRevA.58.400.

\bibitem{Kramer2001}
\bibinfo{author}{Kramer, T.}, \bibinfo{author}{Bracher, C.} \&
  \bibinfo{author}{Kleber, M.}
\newblock \bibinfo{title}{Four-path interference and uncertainty principle in
  photodetachment microscopy}.
\newblock \emph{\bibinfo{journal}{Europhysics Letters (EPL)}}
  \textbf{\bibinfo{volume}{56}}, \bibinfo{pages}{471--477}
  (\bibinfo{year}{2001}).
\newblock \doiprefix 10.1209/epl/i2001-00543-x.

\bibitem{Bracher2005}
\bibinfo{author}{Bracher, C.}, \bibinfo{author}{Delos, J.~B.},
  \bibinfo{author}{Kanellopoulos, V.}, \bibinfo{author}{Kleber, M.} \&
  \bibinfo{author}{Kramer, T.}
\newblock \bibinfo{title}{The photoelectric effect in external fields}.
\newblock \emph{\bibinfo{journal}{Physics Letters A}}
  \textbf{\bibinfo{volume}{347}}, \bibinfo{pages}{62--66}
  (\bibinfo{year}{2005}).
\newblock \doiprefix 10.1016/j.physleta.2005.06.107.

\bibitem{Kanellopoulos2009}
\bibinfo{author}{Kanellopoulos, V.}, \bibinfo{author}{Kleber, M.} \&
  \bibinfo{author}{Kramer, T.}
\newblock \bibinfo{title}{Use of {{Lambert}}'s theorem for the n-dimensional
  {{Coulomb}} problem}.
\newblock \emph{\bibinfo{journal}{Physical Review A}}
  \textbf{\bibinfo{volume}{80}} (\bibinfo{year}{2009}).
\newblock \doiprefix 10.1103/PhysRevA.80.012101.

\bibitem{Kramer2011a}
\bibinfo{author}{Kramer, T.}
\newblock \bibinfo{title}{Time-dependent approach to transport and scattering
  in atomic and mesoscopic physics}.
\newblock \emph{\bibinfo{journal}{AIP Conf. Proc.}}
  \textbf{\bibinfo{volume}{1334}}, \bibinfo{pages}{142--165}
  (\bibinfo{year}{2011}).
\newblock \doiprefix 10.1063/1.3555481.

\bibitem{Blondel1996}
\bibinfo{author}{Blondel, C.}, \bibinfo{author}{Delsart, C.} \&
  \bibinfo{author}{Dulieu, F.}
\newblock \bibinfo{title}{The {{Photodetachment Microscope}}}.
\newblock \emph{\bibinfo{journal}{Physical Review Letters}}
  \textbf{\bibinfo{volume}{77}}, \bibinfo{pages}{3755--3758}
  (\bibinfo{year}{1996}).
\newblock \doiprefix 10.1103/PhysRevLett.77.3755.

\bibitem{Blondel1999}
\bibinfo{author}{Blondel, C.}, \bibinfo{author}{Delsart, C.},
  \bibinfo{author}{Dulieu, F.} \& \bibinfo{author}{Valli, C.}
\newblock \bibinfo{title}{Photodetachment microscopy of {{O}}-}.
\newblock \emph{\bibinfo{journal}{The European Physical Journal D}}
  \textbf{\bibinfo{volume}{5}}, \bibinfo{pages}{207} (\bibinfo{year}{1999}).

\bibitem{Greene2000}
\bibinfo{author}{Greene, C.~H.}, \bibinfo{author}{Dickinson, A.~S.} \&
  \bibinfo{author}{Sadeghpour, H.~R.}
\newblock \bibinfo{title}{Creation of {{Polar}} and {{Nonpolar
  Ultra}}-{{Long}}-{{Range Rydberg Molecules}}}.
\newblock \emph{\bibinfo{journal}{Physical Review Letters}}
  \textbf{\bibinfo{volume}{85}}, \bibinfo{pages}{2458--2461}
  (\bibinfo{year}{2000}).
\newblock \doiprefix 10.1103/PhysRevLett.85.2458.

\bibitem{Bendkowsky2009}
\bibinfo{author}{Bendkowsky, V.} \emph{et~al.}
\newblock \bibinfo{title}{Observation of ultralong-range {{Rydberg}}
  molecules}.
\newblock \emph{\bibinfo{journal}{Nature}} \textbf{\bibinfo{volume}{458}},
  \bibinfo{pages}{1005--1008} (\bibinfo{year}{2009}).
\newblock \doiprefix 10.1038/nature07945.

\bibitem{Granger2001}
\bibinfo{author}{Granger, B.}, \bibinfo{author}{Hamilton, E.} \&
  \bibinfo{author}{Greene, C.}
\newblock \bibinfo{title}{Quantum and semiclassical analysis of long-range
  {{Rydberg}} molecules}.
\newblock \emph{\bibinfo{journal}{Physical Review A}}
  \textbf{\bibinfo{volume}{64}} (\bibinfo{year}{2001}).
\newblock \doiprefix 10.1103/PhysRevA.64.042508.

\bibitem{Lambert1761}
\bibinfo{author}{Lambert, J.~H.}
\newblock \emph{\bibinfo{title}{{Insigniores orbitae cometarum proprietates.}}}
  (\bibinfo{publisher}{{Klett}}, \bibinfo{address}{{Augusae Vindelicorum}},
  \bibinfo{year}{1761}).

\bibitem{Tannor2008}
\bibinfo{author}{Tannor, D.}
\newblock \emph{\bibinfo{title}{Introduction to Quantum Mechanics: A
  Time-Dependent Perspective.}} (\bibinfo{publisher}{{University Science
  Books}}, \bibinfo{address}{{Sausalito}}, \bibinfo{year}{2008}).

\bibitem{Heller2018}
\bibinfo{author}{Heller, E.~J.}
\newblock \emph{\bibinfo{title}{The Semiclassical Way to Dynamics and
  Spectroscopy}} (\bibinfo{publisher}{{Princeton University Press,}},
  \bibinfo{address}{{Princeton}}, \bibinfo{year}{2018}).

\bibitem{Aidala2007}
\bibinfo{author}{Aidala, K.~E.} \emph{et~al.}
\newblock \bibinfo{title}{Imaging magnetic focusing of coherent electron
  waves}.
\newblock \emph{\bibinfo{journal}{Nature Physics}}
  \textbf{\bibinfo{volume}{3}}, \bibinfo{pages}{464--468}
  (\bibinfo{year}{2007}).
\newblock \doiprefix 10.1038/nphys628.

\bibitem{Kramer2008}
\bibinfo{author}{Kramer, T.}, \bibinfo{author}{Heller, E.~J.} \&
  \bibinfo{author}{Parrott, R.~E.}
\newblock \bibinfo{title}{An efficient and accurate method to obtain the
  energy-dependent {{Green}} function for general potentials}.
\newblock \emph{\bibinfo{journal}{Journal of Physics: Conference Series}}
  \textbf{\bibinfo{volume}{99}}, \bibinfo{pages}{012010}
  (\bibinfo{year}{2008}).
\newblock \doiprefix 10.1088/1742-6596/99/1/012010.

\bibitem{Kramer2010}
\bibinfo{author}{Kramer, T.}, \bibinfo{author}{Kreisbeck, C.} \&
  \bibinfo{author}{Krueckl, V.}
\newblock \bibinfo{title}{Wave packet approach to transport in mesoscopic
  systems}.
\newblock \emph{\bibinfo{journal}{Physica Scripta}}
  \textbf{\bibinfo{volume}{82}}, \bibinfo{pages}{038101}
  (\bibinfo{year}{2010}).
\newblock \doiprefix 10.1088/0031-8949/82/03/038101.

\bibitem{Kramer2016AB}
\bibinfo{author}{Kramer, T.} \emph{et~al.}
\newblock \bibinfo{title}{Thermal energy and charge currents in multi-terminal
  nanorings}.
\newblock \emph{\bibinfo{journal}{AIP Advances}} \textbf{\bibinfo{volume}{6}},
  \bibinfo{pages}{065306} (\bibinfo{year}{2016}).
\newblock \doiprefix 10.1063/1.4953812.

\bibitem{Kreisbeck2017}
\bibinfo{author}{Kreisbeck, C.}, \bibinfo{author}{Kramer, T.} \&
  \bibinfo{author}{Molina, R.~A.}
\newblock \bibinfo{title}{Time-dependent wave packet simulations of transport
  through {{Aharanov}}\textendash{{Bohm}} rings with an embedded quantum dot}.
\newblock \emph{\bibinfo{journal}{Journal of Physics: Condensed Matter}}
  \textbf{\bibinfo{volume}{29}}, \bibinfo{pages}{155301}
  (\bibinfo{year}{2017}).
\newblock \doiprefix 10.1088/1361-648X/aa605d.

\bibitem{Benner1995}
\bibinfo{author}{Benner, L.}
\newblock \bibinfo{title}{On the {{Orbital Evolution}} and {{Origin}} of
  {{Comet Shoemaker}}-{{Levy}} 9}.
\newblock \emph{\bibinfo{journal}{Icarus}} \textbf{\bibinfo{volume}{118}},
  \bibinfo{pages}{155--168} (\bibinfo{year}{1995}).
\newblock \doiprefix 10.1006/icar.1995.1182.

\bibitem{Vercin1991}
\bibinfo{author}{Ver{\c c}in, A.}
\newblock \bibinfo{title}{Two anyons in a static, uniform magnetic field.
  {{Exact}} solution}.
\newblock \emph{\bibinfo{journal}{Physics Letters B}}
  \textbf{\bibinfo{volume}{260}}, \bibinfo{pages}{120--124}
  (\bibinfo{year}{1991}).
\newblock \doiprefix 10.1016/0370-2693(91)90978-Y.

\bibitem{Kramer2010a}
\bibinfo{author}{Kramer, T.}, \bibinfo{author}{Kreisbeck, C.} \&
  \bibinfo{author}{Krueckl, V.}
\newblock \bibinfo{title}{Wave packet approach to transport in mesoscopic
  systems}.
\newblock \emph{\bibinfo{journal}{Physica Scripta}}
  \textbf{\bibinfo{volume}{82}}, \bibinfo{pages}{038101}
  (\bibinfo{year}{2010}).
\newblock \doiprefix 10.1088/0031-8949/82/03/038101.

\bibitem{Grossmann2011}
\bibinfo{author}{Grossmann, F.} \& \bibinfo{author}{Kramer, T.}
\newblock \bibinfo{title}{Spectra of harmonium in a magnetic field using an
  initial value representation of the semiclassical propagator}.
\newblock \emph{\bibinfo{journal}{Journal of Physics A: Mathematical and
  Theoretical}} \textbf{\bibinfo{volume}{44}}, \bibinfo{pages}{445309}
  (\bibinfo{year}{2011}).
\newblock \doiprefix 10.1088/1751-8113/44/44/445309.

\bibitem{Kramer2015}
\bibinfo{author}{Kramer, T.} \& \bibinfo{author}{Noack, M.}
\newblock \bibinfo{title}{Prevailing dust-transport directions on comet
  {{67P}}/{{Churyumov}}\textendash{{Gerasimenko}}}.
\newblock \emph{\bibinfo{journal}{The Astrophysical Journal}}
  \textbf{\bibinfo{volume}{813}}, \bibinfo{pages}{L33--L33}
  (\bibinfo{year}{2015}).
\newblock \doiprefix 10.1088/2041-8205/813/2/L33.

\bibitem{VanHouten1989}
\bibinfo{author}{{van Houten}, H.} \emph{et~al.}
\newblock \bibinfo{title}{Coherent electron focusing with quantum point
  contacts in a two-dimensional electron gas}.
\newblock \emph{\bibinfo{journal}{Physical Review B}}
  \textbf{\bibinfo{volume}{39}}, \bibinfo{pages}{8556--8575}
  (\bibinfo{year}{1989}).
\newblock \doiprefix 10.1103/PhysRevB.39.8556.

\bibitem{Aoki1981}
\bibinfo{author}{Aoki, T.} \& \bibinfo{author}{{Ando, T.}}
\newblock \bibinfo{title}{Effect of localization on the {{Hall}} conductivity
  in the two-dimensional system in strong magnetic fields}.
\newblock \emph{\bibinfo{journal}{Solid State Communications}}
  \textbf{\bibinfo{volume}{38}}, \bibinfo{pages}{1079--1082}
  (\bibinfo{year}{1981}).
\newblock \doiprefix 10.1016/0038-1098(81)90021-1.

\bibitem{Prange1981}
\bibinfo{author}{Prange, R.}
\newblock \bibinfo{title}{Quantized {{Hall}} resistance and the measurement of
  the fine-structure constant}.
\newblock \emph{\bibinfo{journal}{Physical Review B}}
  \textbf{\bibinfo{volume}{23}}, \bibinfo{pages}{4802} (\bibinfo{year}{1981}).
\newblock \doiprefix 10.1103/PhysRevB.23.4802.

\bibitem{Uemura1983}
\bibinfo{editor}{Uemura, Y.}, \bibinfo{editor}{Kamimura, H.} \&
  \bibinfo{editor}{Toyozawa, Y.} (eds.) \emph{\bibinfo{title}{Recent Topics in
  Semiconductor Physics: In Commemoration of the Sixtieth Birthday of
  {{Yasutada Uemura}}}} (\bibinfo{publisher}{{World Scientific}},
  \bibinfo{address}{{Singapore}}, \bibinfo{year}{1983}).

\bibitem{Brenig1983}
\bibinfo{author}{Brenig, W.}
\newblock \bibinfo{title}{Quantized {{Hall}} conductance: {{Effect}} of random
  potentials}.
\newblock \emph{\bibinfo{journal}{Zeitschrift f\"ur Physik B Condensed Matterur
  Physik B Condensed Matter}} \textbf{\bibinfo{volume}{50}},
  \bibinfo{pages}{305--309} (\bibinfo{year}{1983}).
\newblock \doiprefix 10.1007/BF01470042.

\bibitem{Beenakker1991}
\bibinfo{author}{Beenakker, C. W.~J.} \& \bibinfo{author}{{van Houten}, H.}
\newblock \bibinfo{title}{Quantum {{Transport}} in {{Semiconductor
  Nanostructures}}}.
\newblock In \bibinfo{editor}{Ehrenreich, H.} \& \bibinfo{editor}{Turnbull, D.}
  (eds.) \emph{\bibinfo{booktitle}{Solid {{State Physics}}}},
  vol.~\bibinfo{volume}{44} of \emph{\bibinfo{series}{Semiconductor
  {{Heterostructures}} and {{Nanostructures}}}}, \bibinfo{pages}{1--228}
  (\bibinfo{publisher}{{Academic Press}}, \bibinfo{year}{1991}).

\bibitem{Johnson1983}
\bibinfo{author}{Johnson, B.~R.}, \bibinfo{author}{Hirschfelder, J.~O.} \&
  \bibinfo{author}{Yang, K.-H.}
\newblock \bibinfo{title}{Interaction of atoms, molecules, and ions with
  constant electric and magnetic fields}.
\newblock \emph{\bibinfo{journal}{Reviews of Modern Physics}}
  \textbf{\bibinfo{volume}{55}}, \bibinfo{pages}{109--153}
  (\bibinfo{year}{1983}).
\newblock \doiprefix 10.1103/RevModPhys.55.109.

\bibitem{Ehrenfest1927}
\bibinfo{author}{Ehrenfest, P.}
\newblock \bibinfo{title}{{Bemerkung \"uber die angen\"aherte G\"ultigkeit der
  klassischen Mechanik innerhalb der Quantenmechanik}}.
\newblock \emph{\bibinfo{journal}{Zeitschrift f\"ur Physik}}
  \textbf{\bibinfo{volume}{45}}, \bibinfo{pages}{455--457}
  (\bibinfo{year}{1927}).
\newblock \doiprefix 10.1007/BF01329203.

\bibitem{Berry1972}
\bibinfo{author}{Berry, M.~V.} \& \bibinfo{author}{Mount, K.~E.}
\newblock \bibinfo{title}{Semiclassical approximations in wave mechanics}.
\newblock \emph{\bibinfo{journal}{Reports on Progress in Physics}}
  \textbf{\bibinfo{volume}{35}}, \bibinfo{pages}{315--397}
  (\bibinfo{year}{1972}).
\newblock \doiprefix 10.1088/0034-4885/35/1/306.

\bibitem{Kramer2005}
\bibinfo{author}{Kramer, T.} \& \bibinfo{author}{Bracher, C.}
\newblock \bibinfo{title}{Propagation in {{Crossed Electric}} and {{Magnetic
  Fields}}: {{The Quantum Source Approach}}}.
\newblock In \bibinfo{editor}{Gruber, B.~J.}, \bibinfo{editor}{Marmo, G.} \&
  \bibinfo{editor}{Yoshinaga, N.} (eds.) \emph{\bibinfo{booktitle}{Symmetries
  in {{Science XI}}}}, \bibinfo{pages}{317--353} (\bibinfo{publisher}{{Kluwer
  Academic Publishers}}, \bibinfo{address}{{Dordrecht}}, \bibinfo{year}{2005}).

\bibitem{Grosche2013}
\bibinfo{author}{Grosche, C.}
\newblock \emph{\bibinfo{title}{Handbook of Feynman Path Integrals.}}
  (\bibinfo{publisher}{{Springer}}, \bibinfo{address}{{Berlin}},
  \bibinfo{year}{2013}).

\bibitem{Berry1981}
\bibinfo{author}{Berry, M.~V.}
\newblock \bibinfo{title}{A curious multifoliate caustic in the magnetic
  {{Green}} function}.
\newblock \emph{\bibinfo{journal}{European Journal of Physics}}
  \textbf{\bibinfo{volume}{2}}, \bibinfo{pages}{22--28} (\bibinfo{year}{1981}).
\newblock \doiprefix 10.1088/0143-0807/2/1/003.

\bibitem{Kramer2004}
\bibinfo{author}{Kramer, T.}, \bibinfo{author}{Bracher, C.} \&
  \bibinfo{author}{Kleber, M.}
\newblock \bibinfo{title}{Electron propagation in crossed magnetic and electric
  fields}.
\newblock \emph{\bibinfo{journal}{Journal of Optics B: Quantum and
  Semiclassical Optics}} \textbf{\bibinfo{volume}{6}}, \bibinfo{pages}{21--27}
  (\bibinfo{year}{2004}).
\newblock \doiprefix 10.1088/1464-4266/6/1/004.

\bibitem{Kramer2005c}
\bibinfo{author}{Kramer, T.}
\newblock \bibinfo{title}{Electron drift orbits in crossed electromagnetic
  fields and the quantum {{Hall}} effect}.
\newblock In \emph{\bibinfo{booktitle}{Inst. {{Phys}}. {{Conf}}. {{Series}}}},
  vol. \bibinfo{volume}{185}, \bibinfo{pages}{353} (\bibinfo{publisher}{{CRC
  Press}}, \bibinfo{address}{{Cocoyoc, Mexico}}, \bibinfo{year}{2005}).
\newblock \eprint{cond-mat/0512690}.

\bibitem{Davies1998}
\bibinfo{author}{Davies, J.~H.}
\newblock \emph{\bibinfo{title}{The Physics of Low-Dimensional Semiconductors
  an Introduction}} (\bibinfo{publisher}{{Cambridge University Press}},
  \bibinfo{address}{{Cambridge}}, \bibinfo{year}{1998}).

\bibitem{Kramer2010b}
\bibinfo{author}{Kramer, T.}
\newblock \bibinfo{title}{Two interacting electrons in a magnetic field:
  Comparison of semiclassical, quantum, and variational solutions}.
\newblock \emph{\bibinfo{journal}{AIP Conference Proceedings}}
  \textbf{\bibinfo{volume}{1323}}, \bibinfo{pages}{178--190}
  (\bibinfo{year}{2010}).
\newblock \doiprefix 10.1063/1.3537846.

\bibitem{Bateman1953}
\bibinfo{author}{Bateman, H.}
\newblock \emph{\bibinfo{title}{Higher {{Transcendental Functions}} [{{Volumes
  I}}-{{III}}]}}, vol. \bibinfo{volume}{I-III}
  (\bibinfo{publisher}{{McGraw-Hill Book Company}}, \bibinfo{address}{{New
  York}}, \bibinfo{year}{1953}).

\bibitem{Bracher2006}
\bibinfo{author}{Bracher, C.}, \bibinfo{author}{Kramer, T.} \&
  \bibinfo{author}{Delos, J.~B.}
\newblock \bibinfo{title}{Electron dynamics in parallel electric and magnetic
  fields}.
\newblock \emph{\bibinfo{journal}{Physical Review A}}
  \textbf{\bibinfo{volume}{73}} (\bibinfo{year}{2006}).
\newblock \doiprefix 10.1103/PhysRevA.73.062114.

\bibitem{Broucke1968}
\bibinfo{author}{Broucke, R.~A.}
\newblock \bibinfo{title}{Periodic {{Orbits}} in the {{Restricted
  Three}}-{{Body Problem With Earth}}-{{Moon Masses}}}.
\newblock \emph{\bibinfo{journal}{JPL NASA Technical Report}}
  \textbf{\bibinfo{volume}{32-1168}}, \bibinfo{pages}{100}
  (\bibinfo{year}{1968}).

\bibitem{Anderson2019}
\bibinfo{author}{Anderson, R.~L.}, \bibinfo{author}{Chodas, P.~W.},
  \bibinfo{author}{Easton, R.~W.} \& \bibinfo{author}{Lo, M.~W.}
\newblock \bibinfo{title}{Planar low-energy asteroid and comet transit analysis
  using isolating blocks}.
\newblock \emph{\bibinfo{journal}{Celestial Mechanics and Dynamical Astronomy}}
  \textbf{\bibinfo{volume}{131}} (\bibinfo{year}{2019}).
\newblock \doiprefix 10.1007/s10569-019-9909-1.

\bibitem{Friedrich1989}
\bibinfo{author}{Friedrich, H.} \& \bibinfo{author}{Wintgen, D.}
\newblock \bibinfo{title}{The hydrogen atom in a uniform magnetic field -
  {{An}} example of chaos}.
\newblock \emph{\bibinfo{journal}{Physics Reports}}
  \textbf{\bibinfo{volume}{183}}, \bibinfo{pages}{37--79}
  (\bibinfo{year}{1989}).
\newblock \doiprefix 10.1016/0370-1573(89)90121-X.

\bibitem{Bessel1836}
\bibinfo{author}{Bessel, F., Geheimer Rat und~Ritter}.
\newblock \bibinfo{title}{{Bemerkungen {\"u}ber m{\"o}gliche
  Unzul{\"a}nglichkeit der die Anziehungen allein ber{\"u}cksichtigenden
  Theorie der Kometen}}.
\newblock \emph{\bibinfo{journal}{Astronomische Nachrichten}}
  \textbf{\bibinfo{volume}{13}}, \bibinfo{pages}{345--350}
  (\bibinfo{year}{1836}).
\newblock \doiprefix 10.1002/asna.18360132302.

\bibitem{Kramer2017}
\bibinfo{author}{Kramer, T.}, \bibinfo{author}{L{\"a}uter, M.},
  \bibinfo{author}{Rubin, M.} \& \bibinfo{author}{Altwegg, K.}
\newblock \bibinfo{title}{Seasonal changes of the volatile density in the coma
  and on the surface of comet {{67P}}/{{Churyumov}}-{{Gerasimenko}}}.
\newblock \emph{\bibinfo{journal}{Monthly Notices of the Royal Astronomical
  Society}} \textbf{\bibinfo{volume}{469}}, \bibinfo{pages}{S20--S28}
  (\bibinfo{year}{2017}).
\newblock \doiprefix 10.1093/mnras/stx866.

\bibitem{Lauter2019}
\bibinfo{author}{L{\"a}uter, M.}, \bibinfo{author}{Kramer, T.},
  \bibinfo{author}{Rubin, M.} \& \bibinfo{author}{Altwegg, K.}
\newblock \bibinfo{title}{Surface localization of gas sources on comet
  {{67P}}/{{Churyumov}}-{{Gerasimenko}} based on {{DFMS}}/{{COPS}} data}.
\newblock \emph{\bibinfo{journal}{Monthly Notices of the Royal Astronomical
  Society}} \textbf{\bibinfo{volume}{483}}, \bibinfo{pages}{852--861}
  (\bibinfo{year}{2019}).
\newblock \doiprefix 10.1093/mnras/sty3103.

\bibitem{Kramer2019}
\bibinfo{author}{Kramer, T.} \emph{et~al.}
\newblock \bibinfo{title}{Comet {{67P}}/{{Churyumov}}-{{Gerasimenko}} rotation
  changes derived from sublimation-induced torques}.
\newblock \emph{\bibinfo{journal}{Astronomy \& Astrophysics}}
  \textbf{\bibinfo{volume}{630}}, \bibinfo{pages}{A3} (\bibinfo{year}{2019}).
\newblock \doiprefix 10.1051/0004-6361/201834349.

\bibitem{Kramer2019a}
\bibinfo{author}{Kramer, T.} \& \bibinfo{author}{L{\"a}uter, M.}
\newblock \bibinfo{title}{Outgassing-induced acceleration of comet
  {{67P}}/{{Churyumov}}-{{Gerasimenko}}}.
\newblock \emph{\bibinfo{journal}{Astronomy \& Astrophysics}}
  \textbf{\bibinfo{volume}{630}}, \bibinfo{pages}{A4} (\bibinfo{year}{2019}).
\newblock \doiprefix 10.1051/0004-6361/201935229.

\bibitem{Heller1978}
\bibinfo{author}{Heller, E.~J.}
\newblock \bibinfo{title}{Quantum corrections to classical photodissociation
  models}.
\newblock \emph{\bibinfo{journal}{The Journal of Chemical Physics}}
  \textbf{\bibinfo{volume}{68}}, \bibinfo{pages}{2066--2075}
  (\bibinfo{year}{1978}).
\newblock \doiprefix 10.1063/1.436029.

\bibitem{Heller2019}
\bibinfo{author}{Heller, E.~J.}, \bibinfo{author}{Fleischmann, R.} \&
  \bibinfo{author}{Kramer, T.}
\newblock \bibinfo{title}{Branched {{Flow}}}.
\newblock \emph{\bibinfo{journal}{arXiv:1910.07086 [physics]}}
  (\bibinfo{year}{2019}).
\newblock \eprint{1910.07086}.

\end{thebibliography}
\end{document}